\documentclass[aps,prb,showpacs,superscriptaddress,twocolumn,10pt,floatfix]{revtex4-1}
\usepackage{graphicx}
\usepackage{amsmath, amsthm, amssymb}

\usepackage{microtype}
\usepackage{bm}
\usepackage{hyperref}
\usepackage{color}

\newcommand{\Pf}{\text{Pf}}

\newcommand{\up}{\uparrow}
\newcommand{\down}{\downarrow}
\newcommand{\mbb}[1]{\mathbb{#1}}
\newcommand{\mc}[1]{\mathcal{#1}}

\newcommand{\hc}{{\rm H.c.}}
\newcommand{\avg}[1]{\langle #1\rangle}

\def\e{\epsilon}
\def\la{\langle}
\def\ra{\rangle}
\def\sgn{\textrm{sgn}}

\begin{document}
\title{From fractionally charged solitons to Majorana bound states in a one-dimensional interacting model}
\author{Doru Sticlet}
\email{doru-cristian.sticlet@u-bordeaux1.fr}
\affiliation{LOMA (UMR-5798), CNRS and University Bordeaux 1, F-33045 Talence, France}
\affiliation{Max-Planck-Institut f\"ur Physik komplexer Systeme, N\"othnitzer Str. 38, 01187 Dresden, Germany}

\author{Luis Seabra}
\email{seabra@physics.technion.ac.il}
\affiliation{Max-Planck-Institut f\"ur Physik komplexer Systeme, N\"othnitzer Str. 38, 01187 Dresden, Germany}
\affiliation{Department of Physics, Technion - Israel Institute of Technology, Haifa 32000, Israel}

\author{Frank Pollmann}
\email{frankp@pks.mpg.de}
\affiliation{Max-Planck-Institut f\"ur Physik komplexer Systeme, N\"othnitzer Str. 38, 01187 Dresden, Germany}

\author{J\'er\^ome Cayssol}
\email{jerome.cayssol@u-bordeaux1.fr}
\affiliation{LOMA (UMR-5798), CNRS and University Bordeaux 1, F-33045 Talence, France}
\affiliation{Max-Planck-Institut f\"ur Physik komplexer Systeme, N\"othnitzer Str. 38, 01187 Dresden, Germany}

\begin{abstract}
We consider one-dimensional topological insulators hosting fractionally charged midgap states in the presence and absence of induced superconductivity pairing. Under the protection of a discrete symmetry, relating positive and negative energy states, the solitonic midgap states remain pinned at zero energy when superconducting correlations are induced by proximity effect. When the superconducting pairing dominates the initial insulating gap, Majorana fermion phases develop for a class of insulators. As a concrete example, we study the Creutz model with induced $s$-wave superconductivity and repulsive Hubbard-type interactions.
For a finite wire, without interactions, the solitonic modes originating from the nonsuperconducting model survive at zero energy, revealing a fourfold-degenerate ground state. 
However, interactions break the aforementioned discrete symmetry and completely remove this degeneracy, thereby producing a unique ground state which is characterized by a topological bulk invariant with respect to the product of fermion parity and bond inversion. In contrast, the Majorana edge modes are globally robust to interactions. Moreover, the parameter range for which a topological Majorana phase is stabilized expands when increasing the repulsive Hubbard interaction. The topological phase diagram of the interacting model is obtained using a combination of mean-field theory and density matrix renormalization group techniques.
\end{abstract}
\maketitle
\section{Introduction}

In the context of quantum field theory, Jackiw and Rebbi \cite{Jackiw1976} introduced a general mechanism to generate zero modes with fractional charges. They considered a one-dimensional massless fermion coupled to a bosonic scalar field. Owing to $\mbb Z_2$ symmetry breaking, the bosonic field can acquire a finite uniform expectation value $m(x)=\pm m$, which provides a mass to the fermion. When the scalar field has a kink $m(x)$ interpolating between two opposite values of the mass, the one-particle Dirac equation exhibits a nondegenerate solution which is isolated in the middle of the gap (at zero energy $E=0$), and spatially localized around the mass inversion point $x_c$ where $m(x_c)$ vanishes. This nondegenerate solution is protected by a unitary symmetry that connects each single-electron state $\psi_E$ with energy $E$ to its partner $\psi_{-E}$ located at the opposite energy. The zero-energy state $\psi_{E=0}$ is self-conjugated under this discrete symmetry and thereby protected. In the many-particle description, there are two degenerate many-body ground states corresponding to the state $\psi_{E=0}$ being empty or filled, while negative energy states of the Fermi sea are all filled. Besides, the existence of a U(1) symmetry ensures that the electric charge is a good quantum number. Then, the ground state with $E=0$ empty (occupied) carries the charge $Q=1/2$ ($Q=-1/2$), in units of the original fermion charge. This has been generalized to arbitrary fractional charges \cite{Su1981, Goldstone1981} and more complicated bosonic kinks.\cite{Jackiw1981,Volovik2003} The first condensed matter realization of the Jackiw-Rebbi mechanism came with the study of conducting polymers, such as polyacetylene. \cite{Su1979, Heeger1988}

More recently, another type of zero-energy excitation, the Majorana fermion, has attracted a lot of attention from the condensed matter community.\cite{Alicea2012, *Beenakker2013} A Majorana fermion is its own antiparticle,\cite{Majorana1937} and can appear at defects of topological superconductors, such as $p$-wave superconductors\cite{Read2000} or superconducting hybrid systems mimicking them.\cite{Fu2008,Lutchyn2010,Oreg2010} In presence of superconductivity, the U(1) symmetry is broken, such that the electrical charge is no longer conserved. Moreover, a discrete antiunitary particle-hole symmetry emerges which differs from the one discussed above in the case of fractionally charged solitons. Nevertheless, a Majorana excitation can also be described within the Jackiw-Rebbi scheme and its existence is signaled by a nondegenerate and self-conjugated solution of the Bogoliubov--de Gennes (BdG) equations. \cite{Read2000} The crucial difference appears in the many-particle description of the system: two solutions of the BdG equations are needed to construct a fermionic doublet.  

This paper aims at understanding the physics of systems that host fractionally charged solitons in their normal state and Majorana modes in some of their superconducting phases. In particular, we study how the zero-energy solitonic modes survive under the addition of proximity-induced superconductivity and/or Hubbard repulsive interactions. Several phases are {\it a priori} possible, including a fully gapped superconducting phase without midgap states, a topological superconductor with Majorana end states, and finally a superconducting phase with midgap states which are not Majorana states.
 The general phase diagram of such a system is expected to be ruled by at least two symmetries: the chiral symmetry of the normal system, and the particle-hole symmetry associated with superconductivity. Although the U(1) symmetry is broken, zero-energy states can still be protected by the discrete symmetry between the positive and negative energies. The Majorana bound states (MBS) are end states described by particle-hole self-conjugated wave functions. The solitonic modes which do not have this Majorana self-conjugation property (and eventually survive in the superconducting phase) are henceforth called, interchangeably, chiral bound states (CBS) or, simply, solitons. Recently, the coexistence of CBS and MBS was studied in Rashba nanowires where the localized end states are realized by applying a non uniform magnetic field.\cite{Klinovaja2012a} In this nanowire system, the end states occur usually at finite energy and a fine tuning of the parameters of the model is required to bring these states to zero energy. The BDI symmetry class\cite{Altland1997} can provide a family of models that exhibit both CBS and MBS, over a wide range of their phase diagram, without fine tuning. Indeed, the BDI class harbors both topological insulators and superconductors associated with integer topological numbers ($\mbb Z$).\cite{Schnyder2008,Kitaev2001}  On one hand, this implies that BDI superconductors host multiple Majorana particles which are spatially close but do not hybridize, thereby remaining at zero energy.\cite{Niu2012, Tewari2012,Sticlet2013a} On the other hand, the BDI class also encompasses insulating (nonsuperconducting) systems with zero-energy fractionally charged solitonic end states. The most famous representative example of such BDI nontrivial insulators is the polyacetylene chain described by the Su-Schrieffer-Heeger (SSH) model.\cite{Su1979}

Among the possible BDI systems, this paper investigates in detail the Creutz model,\cite{Creutz1994, Creutz1999} and also some aspects of the SSH model\cite{Su1979} in the Appendix. Initially introduced in the context of lattice quantum chromodynamics, the Creutz model has gained a foothold in condensed matter physics as a versatile model to test different ideas, such as the topological production of defects when crossing a quantum transition point,\cite{Bermudez2009} the dynamics of Dirac points under interaction,\cite{Dora2013} the decay of the edge states in the presence of a thermal bath,\cite{Viyuela2012} or the persistent currents in Dirac fermion rings.\cite{Sticlet2013b} Proposals for the realization of the Creutz model in cold-atom systems have been recently advanced.\cite{Mazza2012} Moreover, phases of the \textit{bosonic} Creutz model were also recently investigated.\cite{Takayoshi2013,Tovmasyan2013}

In the absence of superconductivity, the Creutz model exhibits gapped phases with and without solitonic modes.\cite{Creutz1994, Creutz1999} The transition between these inequivalent gapped phases occurs by closing the bulk gap, thereby leading to a semi-metallic phase with a single Dirac cone at the transition. Our purpose is to enrich the Creutz system by adding spin-singlet superconducting pairing, induced by a proximity effect. 
When the superconducting order parameter becomes larger than the initial insulating gap, a topological superconductor (TSC) phase develops, hosting  Majorana edge modes.
In view of this, the superconducting Creutz model is  henceforth called the Creutz-Majorana (CM) model.
This phenomenology is also derived in the SSH model in the presence of proximity-induced superconducting pairing.
Moreover, the superconducting phase in which the solitons survive proves to be in the bulk a topological phase protected by spatial inversion at a bond and fermion parity. It is topologically distinct from the Majorana phase and from the trivial phase without in-gap states.

The second thread followed by this paper concerns the effect of the repulsive Hubbard interaction on the different phases of the CM model and on its edge states.
The effect of Coulomb  interactions on MBS has been investigated recently in nanowires with strong spin-orbit in the presence of superconducting proximity effect which are relevant for experiments.\cite{Mourik2012} Such nanowire models \cite{Oreg2010, Lutchyn2010}  break the time-reversal symmetry, but they nevertheless map onto a BDI system at low energy, the Kitaev Majorana chain,\cite{Kitaev2001} and it was shown theoretically that Majorana bound states are robust with respect to repulsive Hubbard interactions.\cite{Gangadharaiah2011a}

Soon after, Stoudenmire \textit{et al.}~\cite{Stoudenmire2011} established that MBS are not only robust to interactions, but that, counterintuitively, the parameter regime where the MBS can be observed increases, at the price of a reduction in the bulk superconducting gap.  Following the methodology of Ref.~\onlinecite{Stoudenmire2011}, we study the effect of a repulsive Hubbard interaction on the Creutz-Majorana model using a combination of mean-field theory and  density matrix renormalization group (DMRG). Globally, we obtain  comparable results for the Majorana phase of the Creutz model: its parameter range grows upon increasing the Hubbard interaction. Nevertheless, there is a region in parameter space, where interactions prove detrimental to the Majorana states (cf. Fig.~\ref{fig:MFDMRG}).
In contrast to nanowire models, the models studied in this paper (Creutz and SSH models) host CBS in their normal state in a wide range of parameters. The Hubbard interaction breaks a chiral symmetry that protected the CBS modes, pushing them to finite energy, while the MBS modes remain pinned at zero energy. Even at finite interactions, the phases surrounding the TSC have a different topological character from each other. We describe these phases in terms of their entanglement properties and excitations.

Finally, we notice that recent papers discuss the possibility of having time-reversal-invariant topological superconductors supporting two MBS at each end of a one-dimensional (1D) system.\cite{Zhang2013,Keselman2013,Klinovaja2013} These Kramers pairs of MBS are protected by  coexisting time-reversal symmetry and chiral symmetry. We emphasize that the Creutz model belongs to a completely different symmetry class (BDI rather than DIII) and has at best one MBS at each end. Nevertheless, there are also two symmetries at play, the pseudo-time-reversal symmetry and the chiral symmetry, that allow us to understand the behavior of the solitonic edge modes and of the MBS.

In brief, this paper investigates the mechanism of competition between fractionally charged solitons and Majorana modes in the large class of BDI models, thereby complementing the recent proposal in Ref.~\onlinecite{Klinovaja2013}. Furthermore, we demonstrate that the stabilization of Majoranas by  repulsive interactions, first obtained in Ref.~\onlinecite{Stoudenmire2011}, is also globally observed in a distinct topological insulator, the Creutz model, although its phase diagram also presents regions where the Majoranas are destabilized.

The paper is organized as follows: Section~\ref{sec:C} reviews the Creutz model and its topological properties. Section~\ref{sec:CM} investigates the Creutz-Majorana model, including the survival of the CBS in the superconducting region, and their transition to MBS upon increasing the pairing. Section~\ref{sec:CMH} takes into account the effects of repulsive interactions in the Creutz-Majorana-Hubbard model (CMH) using self-consistent Hartree-Fock approximation and density matrix renormalization group (DMRG). Section~\ref{sec:CH} examines a particular case of the interacting, but nonsuperconducting, Creutz model where a bulk metallic phase develops. Section~\ref{sec:conc} holds the conclusions. The Appendix comments on the generality of the results by treating the noninteracting Su-Schrieffer-Heeger (SSH) polyacetylene model\cite{Su1979} enriched with a $p$-wave superconducting pairing.

\section{Creutz model}
\label{sec:C}
Here, we present the noninteracting part of the lattice model. Historically, this type of model was introduced by Wilson\cite{Wilson1977} to solve the fermion doubling problem and simulate Dirac fermions on lattices. Later, Creutz investigated the existence of edge states in finite chains or ladders.\cite{Creutz1994} We shall refer to this model simply as the Creutz model in the remainder of this paper. This section provides a review of the spectral and topological properties of the Creutz model.

\subsection{Model}

\begin{figure}[t]
\centering
\begin{minipage}[c]{0.39\columnwidth}
\includegraphics[width=\textwidth]{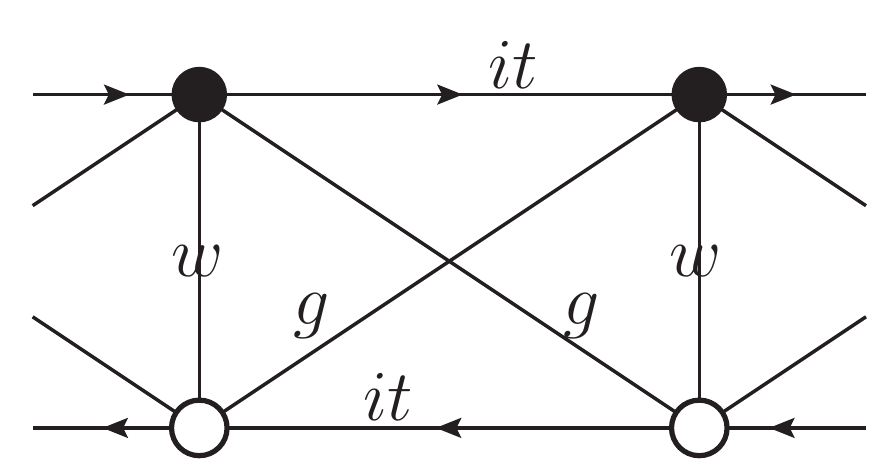}
\end{minipage}
\begin{minipage}[c]{0.59\columnwidth}
\includegraphics[width=\textwidth]{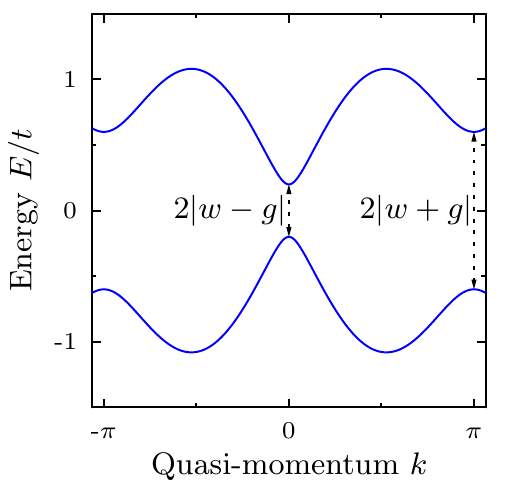}
\end{minipage}
\caption{(Color online) Left: Creutz lattice model. The filled (empty) bullets represent spin-up (-down) states of $\sigma_3$. Each site encompasses a vertical bond, containing two spin states. The spin degeneracy is lifted by an on-site coupling $w$. There is a spin-conserving hopping term $t$, and a ``spin-orbit'' term $g$ that encodes the probability of electrons to spin flip during hopping to a different site. Right: Energy dispersion of the lattice Creutz model. One-dimensional Dirac cones can be realized at $k=0$ and $\pi$ for a parameter choice $w=\pm g$. Otherwise, the system is insulating with gapped Dirac cones as shown in the figure.}
\label{fig:Creutz}
\end{figure}

In the absence of superconductivity and interactions, our starting point is the lattice Hamiltonian
\begin{equation}\label{HCreutz}
H_{C}=\frac{1}{2}\sum_j\big[
wc_j^\dag\sigma_1 c_j+
c^\dag_j(it\sigma_3-g\sigma_1)c_{j-1}\big]+\hc,
\end{equation}
where the sum runs over all sites indexed by $j$ (see illustration in Fig.~\ref{fig:Creutz}). In this paper, the electronic spin is represented by the standard Pauli matrices $\sigma_i$ ($i=1,2,3$). The spin indices for the electron annihilation operators $c_j=(c_{j\up}, c_{j\down})$ and spin matrices are implicit. The electrons can jump from one site to the nearest-neighboring site while conserving their spin: this process has a complex amplitude $\pm it$, i.e., the electrons gain or lose a phase $\pi/2$ when hopping between the same spin states. 
The electrons can also hop between sites  with amplitude $g$ while flipping their spin, which mimics a spin-orbit coupling. 
Finally, there is an onsite mass term $w$ which favors the  polarization of the electronic spin along the $x$ direction. This Hamiltonian can be seen as an adaptation from Ref.~\onlinecite{Creutz1999}, where the chain index of the Creutz ladder has been replaced by the electronic spin. 

The Hamiltonian is diagonalized by Fourier transforming it into momentum space $H_{\rm C}=\sum_kc_k^\dag\mc H_{\rm C}(k)c_k$:
\begin{equation}
\mc H_{\rm C}(k)=t\sin k\sigma_3+(w-g\cos k)\sigma_1.
\end{equation}
Consequently, there are two bands with the energy dispersion
\begin{equation}\label{energyCreutz}
E_\pm=\pm\sqrt{(t\sin k)^2+(w-g\cos k)^2},
\end{equation}
shown in Fig.~\ref{fig:Creutz}.
From Eq.~(\ref{energyCreutz}) it follows that the two energy bands can touch either at $k=0$, or at $k=\pi$. At these momenta, the energy dispersion is linear and there is a Dirac cone band touching. For $g=w$, the band-touching takes place at $k=0$,  and for $g=-w$, at $k=\pi$. When both $g$ and $w$ vanish, the system exhibits two Dirac cones.

When $g$ and $w$ are finite and close in absolute value, the low-energy physics is given at $k=0$ or $k=\pi$, corresponding to one gapped fermion. Without loss of generality, the parameters in the Creutz model ($t,w,g$) can be considered positive. In this case, the Dirac cone forms at $k=0$, and it is gapped with a mass $g-w$, which can be either positive or negative.
If $g/t$ becomes larger than $w/t$, new minima (maxima) for the upper (lower) band in Eq.~(\ref{energyCreutz}) develop near $k=\pm \pi/2$.

\subsection{Topological characterization}
The Creutz model is classified in the BDI class of topological insulators.\cite{Schnyder2008,Kitaev2001} Although it is a normal (nonsuperconducting) system, it is possible to find an antiunitary operator $\sigma_3\mc K$, which anticommutes with the Hamiltonian, and thus realizes a particle-hole symmetry $\sigma_3 \mc H_{\rm C}^*(k)\mc \sigma_3=-\mc H_{\rm C}(-k)$. 
The operator $\mc K$ is the complex-conjugation operator.  
There is also a pseudo-time-reversal symmetry represented by antiunitary operator, $\sigma_1\mc K$, such that $\sigma_1\mc H_{\rm C}^*(k)\sigma_1=\mc H_{\rm C}(-k)$.
Both pseudo-time-reversal and particle-hole symmetry operators square to one.
Finally, there is a chiral symmetry represented by $\sigma_2$, which is proportional to the product $\sigma_1\sigma_3$ of the other two symmetries. The chiral symmetry anticommutes with the Hamiltonian
\begin{equation}
\{\mc H_{\rm C}(k),\sigma_2\}=0,
\end{equation}
thereby allowing us to associate any eigenstate with energy $E$ and momentum $k$ to another eigenstate with opposite energy and the same momentum.

When $g=w$ (resp. $g=-w$), the system is gapless with a Dirac cone at momentum $k=0$ (resp. $k=\pi$). A particular point is the case of vanishing $g$ and $w$, when the system recovers the time-reversal symmetry, and exhibits two Dirac cones at $k=0$ and $\pi$. In all these gapless phases, there is no well-defined global topological winding number.

For all other values of the parameters $(g,w)$, the chain is a fully gapped insulator, whose 
topological properties are encoded in its associated winding number $\mc W$
\begin{equation}\label{winding}
\mc W=\frac{1}{2}\big[\sgn(w+g)-\sgn(w-g)
\big].
\end{equation}
From the expression of the winding number it follows that when $|g/w|<1$, the system is a topologically trivial insulator ($\mc W=0$) and when $|g/w|>1$, it is a topologically nontrivial insulator characterized by $\mc W=\sgn(g)$.\cite{Sticlet2013b}

In the topologically nontrivial state there are chiral zero-energy bound states (CBS)  at the edges of an open system. These localized states are protected by the chiral symmetry $\sigma_2$ and the bulk gap. For a finite chain, the two edge CBS overlap through the insulating bulk. This overlap is exponentially small for chains longer than the spatial extension of the CBS.\cite{Creutz1999}

\section{Creutz-Majorana model}
\label{sec:CM}
A natural question is whether Majorana fermions can be induced in the Creutz model by considering a superconducting pairing potential.
This would not be very surprising because the system has the necessary ingredients to realize Majorana fermions.
Crucially the system has the possibility to realize a single Dirac cone in the Brillouin zone (for $g=\pm w$).
At the Dirac point, the electron wave functions are eigenstates of $\sigma_3$ with zero energy. Then a spin-singlet superconducting pairing could open a gap in the spectrum and Cooper pairs of spin-polarized fermions can be formed.
In this way, $s$-wave pairing plus a topological insulating state could generate a $p$-wave superconducting regime where MBS become possible at the ends of a finite system. 
A similar mechanism was thoroughly investigated in different proposals to realize MBS using the proximity effect to an $s$-wave superconductor.\cite{Fu2008,Lutchyn2010,Oreg2010}

In addition, the Creutz model displays CBS in the inverted gap regime, $g-w>0$ (where all the parameters in the model were considered real positive numbers).
Therefore a key question is whether induced superconductivity immediately removes these modes, and leads to the formation of Majorana fermions. A central finding of this section is that the solitonic (CBS) modes are robust with respect to the superconducting pairing. These modes are protected by the chiral symmetry and they will subsist at finite superconducting pairing. It will be necessary that the bulk gap closes at finite $\Delta$, in order to subsequently have Majorana fermions forming (see Fig.~\ref{fig:phdiag}).

It is noteworthy to remark that the CBS will prove fragile in comparison to the Majorana fermions. The reason is that the CBS crucially depend on the chiral symmetry for staying at zero energy, and they are removed even by scalar potentials that are created by nonmagnetic disorder.
In the strong pairing regime (large $\Delta$, with respect to the other system parameters), it is expected that the topological Majorana phase is destroyed~\cite{Read2000}.

\subsection{Bulk structure}
\subsubsection{Model}
Let us modify the Creutz Hamiltonian~(\ref{HCreutz}) by adding an $s$-wave singlet superconducting pairing with amplitude  $\Delta$. This leads to the Creutz-Majorana model described by the Hamiltonian  
\begin{equation}\label{HCM}
H_{\rm CM}=H_{\rm C}+H_{\Delta},\quad H_{\Delta}=\sum_j\Delta c^\dag_{j\up} c^\dag_{j\down}+\hc ,
\end{equation}
where the sum is taken over all the lattice sites, and the order parameter $\Delta$ can be considered real without any loss of generality.

The momentum-space Hamiltonian is quadratic in the standard BdG form. Let us choose the basis $C_k^\dag=(c^\dag_{k\up},c^\dag_{k\down},c_{-k\up},c_{-k\down})$. In this basis, the Hamiltonian is written as $H_{\rm CM}=\frac{1}{2}\sum_k C^\dag_k\mc H_{\rm CM}(k)C_k$ with
\begin{equation}\label{HCMk}
\mc H_{\rm CM}(k)
=t\sin k\sigma_3\tau_0+(w-g\cos k)\sigma_1\tau_3-\Delta\sigma_2\tau_2.
\end{equation}
The $\sigma$ are the spin Pauli matrices and $\tau$ are Pauli matrices in particle-hole space. The products of two Pauli matrices from different spaces is understood as a tensor product.

By diagonalizing the BdG Hamiltonian $H_{\rm CM}(k)$, it follows that there are four energy bands $\pm E_{\pm}$ satisfying
\begin{equation}\label{bulgSpec}
E_\pm=\sqrt{(t\sin k)^2+(w-g\cos k\pm \Delta)^2}.
\end{equation}
There are four possible gap closings in the system at $k=0$, for $\Delta=\pm(w-g)$, and at $k=\pi$, for $\Delta=\pm(w+g)$. As it will be shown, all these lines mark topological transitions between different gapped topological phases.

\subsubsection{Symmetries}
In the presence of translational invariance, the following symmetries catalog the Creutz-Majorana Hamiltonian in the BDI topological class.\cite{Altland1997,Schnyder2008,Kitaev2001}
There are two symmetries which are represented by anti-unitary operators squaring to one.
First, the Bogoliubov--de Gennes (BdG) Hamiltonian has an intrinsic {\it built-in} particle-hole symmetry 
\begin{equation}\label{PSym}
 \mc P H_{\rm CM}(k) \mc P^{-1} = - H_{\rm CM}(-k)   ,
\end{equation}
represented by $\mc P=\sigma_0\tau_1\mc K $. This anti-unitary operator $\mc P$ anticommutes with the BdG Hamiltonian and squares to one. 

Second, the system also has a pseudo-time-reversal symmetry represented by the  anti-unitary operator $\mc T = \sigma_1\tau_3\mc K$ satisfying 
\begin{equation}
 \mc T H_{\rm CM}(k) \mc T^{-1} = H_{\rm CM}(-k)   ,
\end{equation}
and commuting with the Hamiltonian.

Finally, there is a chiral symmetry proportional to the product of the two symmetries $\mc P$ and $\mc T$:
\begin{equation}
\mc S=\sigma_1\tau_2.
\end{equation}
This symmetry associates to each state with energy $E$, a partner at $-E$, with the same momentum.
These three symmetries are sufficient to place the CM model in the BDI class.

Nevertheless, it is crucial to note a second set of symmetries that also place the model in the same BDI class. There is a second time-reversal symmetry, $\bar{\mc T}^2=1$, and a second chiral symmetry, $\bar{\mc S}$:
\begin{equation}
\bar{\mc T}=\sigma_2\tau_2\mc K,\quad\bar{\mc S}=\sigma_2\tau_3.
\end{equation}
The new chiral symmetry can be considered as inherited from the original Creutz model, since the nonsuperconducting system already exhibits a chiral symmetry represented by $\sigma_2$. Therefore, under the BdG doubling of degrees of freedom, the symmetry can be extended to the hole sector, such that an electron of a definite chirality is reflected in a hole of the same chirality.
Hence, the addition of superconductivity does not destroy this chiral symmetry, but it simply promotes it in the CM model to $\bar{\mc S}=\sigma_2\tau_3$.

Nevertheless, there is an important difference between the two chiral symmetries of the CM model, $S$ and  $\bar{\mc S}$. The latter, $\bar{\mc S}$, remains (as in the Creutz model) fragile in the sense that it can be destroyed by a scalar (nonmagnetic) on-site potential (which multiplies a $\sigma_0\tau_3$ matrix) because $\{\bar{\mc S},\sigma_0\tau_3\}\ne 0$. In contrast, the chiral symmetry $\mc S$ is preserved as it anticommutes with the scalar potential, $\{S,\sigma_0\tau_3\}=0$. Therefore, unless the bulk gap closes or a perturbation breaks the chiral symmetry, it is guaranteed that the zero-energy modes of the topological insulating phase of the Creutz model remain protected in the Creutz-Majorana model.

The fragility of the chiral symmetry to (local or global) changes in the chemical potential will prove central in explaining the phenomenology of the topological phases in the Creutz-Majorana model with and without interactions.

\subsubsection{Phase diagram}
A one-dimensional BdG Hamiltonian from the BDI class always has an associated $\mbb Z$ topological invariant which is a winding number.\cite{Schnyder2008} Using either one of the two chiral symmetries, it is possible to block off-diagonalize the CM model and to obtain
\begin{equation}
\mc H_{\rm CM}(k)=\begin{pmatrix}
0 & \mc Q(k) \\
\mc Q^\dag(k) & 0
\end{pmatrix}.
\end{equation}
The topological invariant can be extracted from the $\mc Q$ matrix by defining a winding number as
\begin{equation}
\bar{\mc W}=\frac{1}{2\pi i}\oint \frac{dz}{z},\quad z=\frac{\det\mc Q}{|\det\mc Q|}.
\end{equation}
This procedure can be further simplified by using both chiral  symmetries to completely off-diagonalize the Hamiltonian. Subsequently, the winding number can be determined algebraically.

In the present case of the Creutz model, the winding number does not exceed 1 in absolute value, so one can expect at most a single MBS at a given edge. This allows us to use a perhaps more transparent determination of the invariant by using the Majorana number $\mc M$.\cite{Kitaev2001}
This invariant is sensitive only to the winding number parity. But since the Creutz-Majorana model does not realize higher winding number phases, $|\bar{\mc W}|\le 1$, $\mc M$ can decide unequivocally whether the system is in a topological trivial or non-trivial state. In the trivial phase, without Majorana states, the invariant is $\mc M=1$, while in the nontrivial phase (topological superconductor), $\mc M=-1$.

\begin{figure}[t]
\includegraphics[width=\columnwidth]{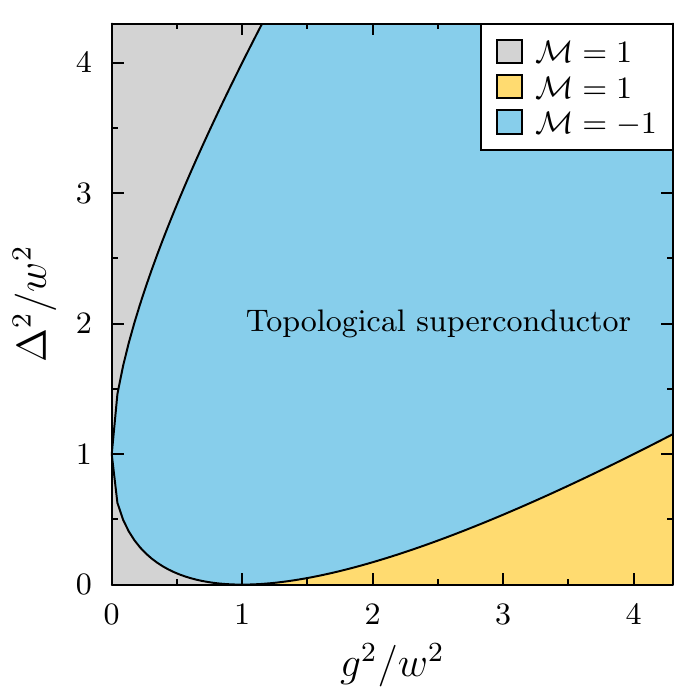}
\caption{(Color online) Phase diagram of the Creutz-Majorana model. The blue region represents the topological superconductor with zero-energy Majorana bound states ($\mc M=-1$). The gray and yellow regions represent phases with $\mc M=1$. The yellow region hosts chiral bound states at zero energy, even if it is trivial with respect to Majorana number $\mc M$.}
\label{fig:phdiag}
\end{figure}

Following Ref.~\onlinecite{Kitaev2001}, the Majorana number is defined at the time-reversal-invariant momenta, $k=0$ and $\pi$
\begin{equation}
\mc M=\sgn\{\Pf[\tau_1\mc H_{\rm CM}(0)]\Pf[\tau_1\mc H_{\rm CM}(\pi)]\},
\end{equation}
where $\Pf$ denotes the Pfaffian of a matrix. It follows immediately that the topological invariant reads as
\begin{equation}\label{MajNo}
\mc M=\sgn\big[\big(1+\frac{g^2}{w^2}-\frac{\Delta^2}{w^2}
\big)^2-4\frac{g^2}{w^2}\big],
\end{equation}
where we have assumed a nonzero on-site $w$.

The ensuing phase diagram is illustrated in two different representations in Figs.~\ref{fig:phdiag} and~\ref{fig:linPhDiag}. Figure~\ref{fig:phdiag} uses Eq.~(\ref{MajNo}) to present the phases as a function of system parameters squared, such that the relative sign difference is not an issue. 
Both the phase without in-gap states (gray) and the CBS hosting phase (yellow) are trivial, with respect to the Majorana number ($\mc M=1$). 
As we shall see in Sec.~\ref{subsec:dmrg}, these two phases are actually distinct with regards to a different topological invariant.
This compact picture can be compared with Fig.~\ref{fig:linPhDiag}, where it is shown more clearly that the topological transition lines in the phase diagram are in fact the lines where the bulk spectrum is gapless [Eq.~(\ref{bulgSpec}]. Figure~\ref{fig:linPhDiag} refines Figs.~\ref{fig:phdiag} by labeling the respective topological phases, and also by illustrating the phase diagram for the original Creutz model ($\Delta/w=0$).

Note in both figures the effect of a finite superconducting pairing $\Delta$ on the original Creutz model, which lies in the horizontal axis $\Delta/w=0$. When  $|g/w|=1$, an infinitesimal $\Delta$ is sufficient to open a superconducting gap at the Dirac cone, and produces a topologically nontrivial superconductor. However, when the Creutz model is deep in a nontrivial insulating state with solitons at its end, the addition of superconducting pairing does not destroy these modes. Since the CBS are protected by the bulk gap and the chiral symmetry $\bar{\mc S}$, they survive the induced superconducting pairing.

Because the solitons develop in a gapped phase without Majorana fermions, the topological invariants $\mc M$ and $\bar{\mc W}$ remain impervious to the existence of the CBS phases. They can not differentiate between trivial phases and the phase with CBS states. As a consequence, the winding number of the Creutz model is not a limit case of the winding number in CM model, $\bar{\mc W}$:
\begin{equation}
\lim_{\Delta\to 0}\bar{\mc W}\ne\mc W.
\end{equation}

Since the chiral bound states' existence can only be inferred from the bulk properties, it is necessary to study their localization on defects in an infinite system, or at the edges of an open system.

\subsection{From chiral bound states to Majorana fermions}
\label{subsec:CtoM}
Here, we further analyze the transition between the trivial superconductor with CBS and the topological superconductor with MBS. Let us first consider a single interface which can host either CBS or MBS. Using a continuum approximation of the lattice model, the single interface problem is solved analytically and the BdG wave functions are derived explicitly. In the second part of this section, we perform numerical diagonalization of a finite length system, which hosts CBS or MBS at its two end points. These studies yield a more precise account of the transition that takes place from two chiral bound states (in the redoubled BdG formalism) to a single Majorana edge mode that develops in the nontrivial superconducting phase.

\subsubsection{Edge state analysis at an interface in a continuum model}
\label{subsec:CBSMBS}

\begin{figure}
\includegraphics[width=\columnwidth]{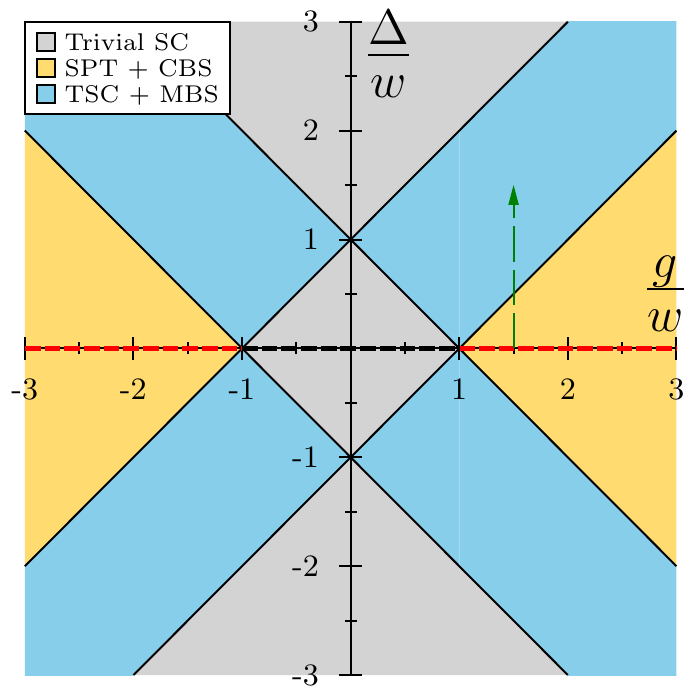}
\caption{(Color online) Phase diagram of the noninteracting superconducting model. The black continuous lines represent the bulk-gap closing. The dashed line on the $x$ axis represents the Creutz model, with CBS at $|g/w|>1$ (in red) and a trivial gapped phase for $|g/w|<1$ (in black). The yellow phases contiguous to the red line represent the domain where the solitonic modes (CBS) survive in the superconducting region. In the bulk, the yellow region is an inversion symmetry-protected topological phase (SPT). The blue region represents the topologically nontrivial superconductor where Majorana bound states can develop.
On the dashed green line, the superconducting pairing increases and produces a transition from the nontrivial insulating Creutz model, to a superconducting phase with CBS, and, subsequently, to the demise of CBS, and the formation of Majorana modes. This transition is investigated in a continuum model in Sec.~\ref{subsec:CBSMBS}.}
\label{fig:linPhDiag}
\end{figure}

Let us consider an isolated interface between a trivial Creutz insulator and the superconducting Creutz-Majorana model (CM). Depending of the topological phase of CM, there will be zero-energy bound states at this interface. Then, by increasing the strength of the superconducting pairing $\Delta$, there is a transition induced from the region with CBS to the one with MBS (see Fig.~\ref{fig:linPhDiag}).  To further fix ideas, let us consider all the parameters in the original models ($t,w,g,\Delta$) as positive real numbers. In this case, the low-energy physics, when $w$ and $g$ are comparable in magnitude, is given by that of a gapped Dirac fermion at $k=0$, with mass $g-w$.
The sign change in the mass signals a transition from a topologically trivial ($g-w<0$) to a nontrivial phase (``inverted'' gap; $g-w>0$) (Eq.~\ref{winding}).

\begin{figure}[t]
\centering
\includegraphics[width=0.7\columnwidth]{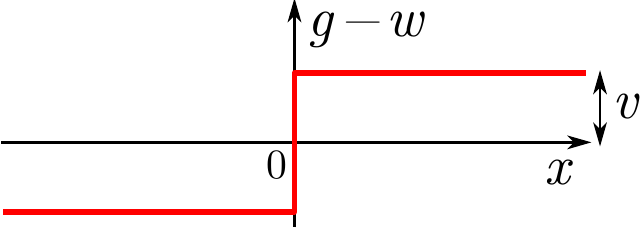}
\caption{Kink in the Dirac fermion mass at the interface. The topologically nontrivial ``inverted'' gap regime is realized for $x>0$.}
\label{fig:mKink}
\end{figure}

Let us start with the Creutz model on an infinite line, and an additional spatial twist is introduced in the mass of the Dirac fermion
\begin{equation}
v(x)=(g-w)\sgn(x)=v\,\sgn(x),
\end{equation}
where $v$ is a positive number, and $x$ is the spatial coordinate (Fig.~\ref{fig:mKink}). According to Eq.~(\ref{winding}), the trivial phase is realized to the left ($x<0$), and the nontrival phase, to the right ($x>0$). At the interface there will be a single solitonic mode (CBS), eigenstate of the chiral symmetry operator $\sigma_2$.

Now, let us investigate the more interesting case of the Creutz-Majorana model, when a superconducting pairing $\Delta$ is added to the nontrivial insulating phase. This will allow us to observe the effect of the superconducting gap on the bound state from the topological insulator. It is now possible to increase the pairing $\Delta$ until the lower boundary to the Majorana phase is crossed (see Fig.~\ref{fig:linPhDiag}). This provides a picture for the behavior of the edge states over the topological transition.

The continuum Hamiltonian that models the interface between the trivial Creutz insulator and the CM model is readily obtained by linearizing the lattice Hamiltonian near $k=0$ and neglecting second-order contributions in momentum
\begin{equation}
\mc H=v_F\sigma_3\tau_0 p-v(x)\sigma_1\tau_3-\theta(x)\Delta\sigma_2\tau_2,
\end{equation}
with $v(x)=v\,\sgn(x)$ and $v=g-w>0$. The function $\theta(x)$ is the Heaviside step function. The Fermi velocity $v_F$ is determined from the lattice model $v_F=ta/\hbar$, with $a$ the lattice constant. 

It is possible to write analytically the wave functions at positive and negative $x$, respectively. The solution is obtained by matching these wave functions at the interface $x=0$. Moreover, the problem is simplified by looking specifically for zero-energy, bound solutions at the interface ($E=0$ and $x=0$).

In the trivial insulating state to the left of the interface, the wave function reads as
\begin{equation}\label{wfLeft}
\psi_L=[\alpha(1,i,0,0)^T+\beta(0,0,1,-i)^T]e^{\frac{vx}{\hbar v_F}},
\end{equation}
where $\alpha$ and $\beta$ are space-independent coefficients, and $T$ transposes the row four-vectors.
In the superconducting phase to the right, the solution will depend on whether it is the superconducting gap or the Dirac fermion mass which dominates. For $\Delta<v$, the right-side wave function reads as
\begin{equation}\label{wfR1}
\psi^{(1)}_R=a_1(1,i,1,-i)^Te^{-\frac{(v+\Delta)x}{\hbar v_F}}+b_1(1,i,-1,i)^Te^{-\frac{(v-\Delta)x}{\hbar v_F}}.
\end{equation}

Matching the wave functions over the boundary, it follows that the coefficients of $\psi_R^{(1)}$ depend on $\alpha$ and $\beta$
\begin{equation}
a_1=(\alpha+\beta)/2,\quad  b_1=(\alpha-\beta)/2.
\end{equation}
Crucially, the matching procedure has left two free coefficients, $\alpha$ and $\beta$. This indicates that there are two modes living at the interface, and they are eigenstates of the chiral symmetry operator $\bar{\mc S}=\sigma_2\tau_3$. These modes are diagonal in electron-hole space, so one of them is electron-like while the other is a hole-like excitation (see Eq.~\ref{wfLeft}). In the doubled (four-band) BdG formalism, they correspond to the solitonic mode that was already present in the Creutz model, in the absence of the superconducting pairing $\Delta$. Breaking the chiral symmetry that protects them allows removal of the CBS from zero energy.
To sum up, when the ``inverted'' Dirac fermion mass dominates the superconducting gap, there are two modes at zero energy protected by the chiral symmetry inherited from the Creutz model.

When the two masses are equal ($\Delta=v$), the bulk gap closes. This signals the topological transition to the Majorana phase. Subsequently, 
for $\Delta>v$, the superconducting gap dominates, and the solution in Eq.~(\ref{wfR1}) becomes non-normalizable at $x>0$. The appropriate (normalizable) wave function in this new regime reads as
\begin{equation}
\psi^{(2)}_R=a_2(1,i,1,-i)^Te^{-\frac{(v+\Delta)x}{\hbar v_F}}
+b_2(-1,i,1,i)^Te^{\frac{(v-\Delta)x}{\hbar v_F}}.
\end{equation}

Matching the right, $\psi^{(2)}_R$, and left, $\psi_L$, wave functions reveals that a single mode remains at the edge in the $\Delta>v$ regime:
\begin{equation}
\psi(x)=\alpha\big[\theta(-x)e^{\frac{vx}{\hbar v_F}}
+\theta(x)e^{-\frac{(v+\Delta)x}{\hbar v_F}}\big](1,i,1,-i)^T.
\end{equation}
At the wave-function level, one notices how the two CBS are combined into a Majorana fermion. Moreover, the resulting wave function has the property that it is self-particle-hole conjugate (for $\alpha\in \mbb R$), indicating that one true Majorana fermion lives now at the interface. As dictated by the bulk-boundary correspondence,\cite{Volovik2003} the edge-state existence could be predicted from information about the bulk topology.

Adding a constant potential $\theta(x)\mu\sigma_0\tau_3$ in the Hamiltonian, removes the CBS from zero energy, but  it does not destroy the Majorana fermion. For a small $\mu$ with respect to both the superconducting gap and the insulating gap, it follows that the domain for Majorana fermions increases, such that the transition from the trivially gapped state to the nontrivial superconductor moves at lower pairing potential $\Delta^2=v^2-\mu^2$. Additionally, if the chemical potential is larger than the insulating gap $\mu>v$, then any positive pairing potential opens a gap which hosts Majorana fermions. This is a case where the potential is in a band and the single-band electrons  are subsequently coupled by a $p$-wave-type pairing to realize a topological superconducting phase. 

In conclusion, we have described the transition between the trivial superconducting phase with doubly-degenerate zero-energy states protected by the chiral symmetry to a phase with Majorana fermions.
This section offers the proof that the local twofold degeneracy in the spectrum does not stand for multiple Majorana modes at an edge, but for chiral bound states protected by a fragile symmetry inherited from the Creutz insulator.

\subsubsection{Spectrum of the finite lattice model: Gapping the CBS}
The CBS in the superconducting phase contiguous to the solitonic phase of the Creutz model, can be equally explored using exact diagonalization of the lattice Hamiltonian in Eq.~(\ref{HCM}). 
In an open wire, the CBS and the MBS form at the two edges of the system. Therefore there will be a doubling in the degeneracy of zero-energy modes.
The numerical analysis of the lattice confirms the location of the bound states, reveals the fourfold energy degeneracy of the CBS, and their sensitivity to chiral symmetry breaking terms.

\begin{figure}[t]
\includegraphics[width=\columnwidth]{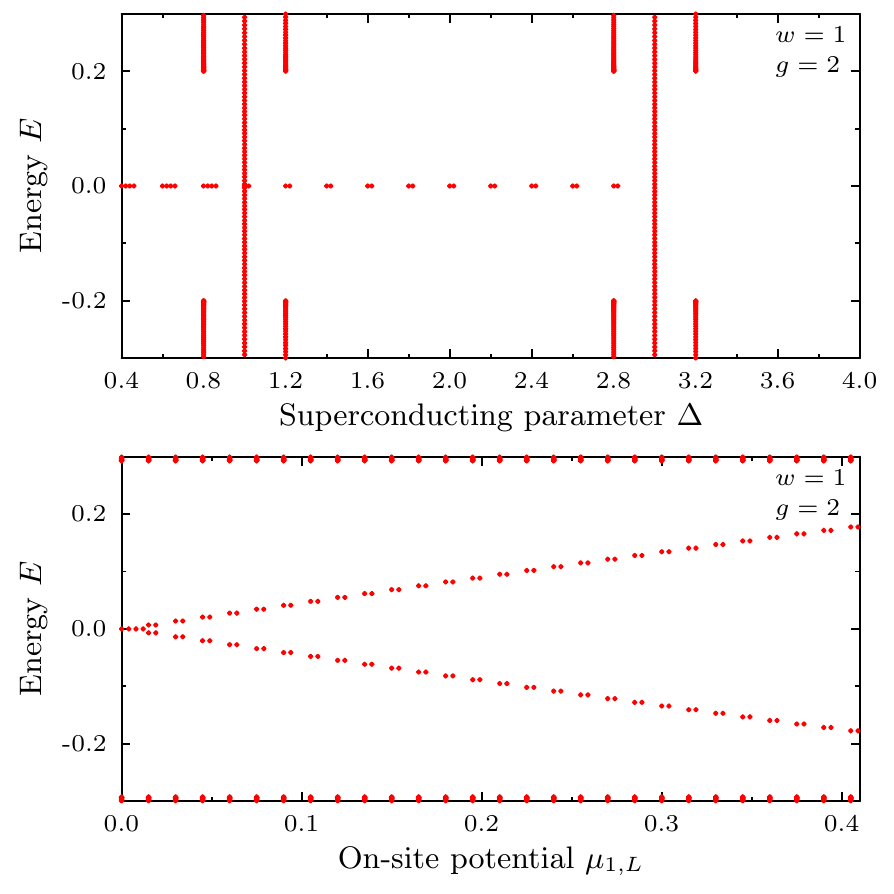}
\caption{Representation of the energy eigenstates of the BdG Hamiltonian near zero energy. Top: Phase transitions at $\Delta/w=1,3$. Majorana modes are present when $\Delta/w\in(1,3)$. Bottom: Removal of the fourfold degeneracy when an onsite scalar potential is added on the first and last sites. The wire length is $L=500 a$.}
\label{fig:degens}
\end{figure}

It is possible to take a crosscut through the phase diagram from Fig.~\ref{fig:linPhDiag}. The system parameters are fixed, except the pairing $\Delta$ which is varied in order to traverse the three distinct regimes of the model: trivial superconductor, superconductor with CBS, and nontrivial superconductor with MBS. In units where $t=1$, the parameters are $g=2$ and $w=1$. The pairing $\Delta$ is varied as to encompass the topological transitions at $\Delta_1=1$ and $\Delta_2=3$. The lowest eigenvalues resulting from exact diagonalization are plotted against $\Delta$ in Fig.~\ref{fig:degens} (top). The chiral bound states are seen as fourfold-degenerate midgap states at small $\Delta$. The bulk spectrum closes at $\Delta_1$ and a twofold degeneracy characteristic of Majorana fermions remains. In the strong-pairing regime $\Delta>\Delta_2$, the system becomes a trivial superconductor devoid of midgap states.

The fragility of the chiral bound states can be tested by adding different types of scalar potentials. Because the chiral symmetry $\bar{\mc S}$ does not anticommute with the local on-site scalar potentials, these modes are not robust.
Nonmagnetic disorder or even a constant chemical potential removes them from zero energy. In fact, it is sufficient to add a local potential on the first and last sites to lift them from zero energy. This chiral-symmetry-breaking term is implemented in Eq.~(\ref{HCM}) by adding
\begin{equation}
\sum_{j=1,L}
C^\dag_j\mu_j\sigma_0\tau_3C_j.
\end{equation}
The result is presented in the spectrum of the lowest eigenstates in Fig.~\ref{fig:degens}(bottom), for $\Delta/w=\sqrt{0.5}$ and $g/w=2$, deep in the CBS region of the phase diagram. Since the chemical potential is chosen identical on the first and last sites $\mu_1=\mu_L$, the fourfold degeneracy is reduced to two degenerate pairs of states moving in opposite energy direction away from zero energy.

In the Creutz model, a chemical potential can move two degenerate states away from zero energy. In the CM superconducting model, a pair of electronic states is reflected in a pair of hole states that feel an opposite chemical potential. This observation accounts for the evolution of states seen in Fig.~\ref{fig:degens} (top). The remaining twofold degeneracy is naturally removed by rendering incommensurate the chemical potential magnitude on the first and last sites. That is the general case of a (nonmagnetically) disordered wire.

In contrast, in the nontrivial superconducting region there is a single Majorana fermion bound at each edge. Moderate disorder which does not close the bulk gap leaves them at zero energy.

\begin{figure}[t]
\includegraphics[width=\columnwidth]{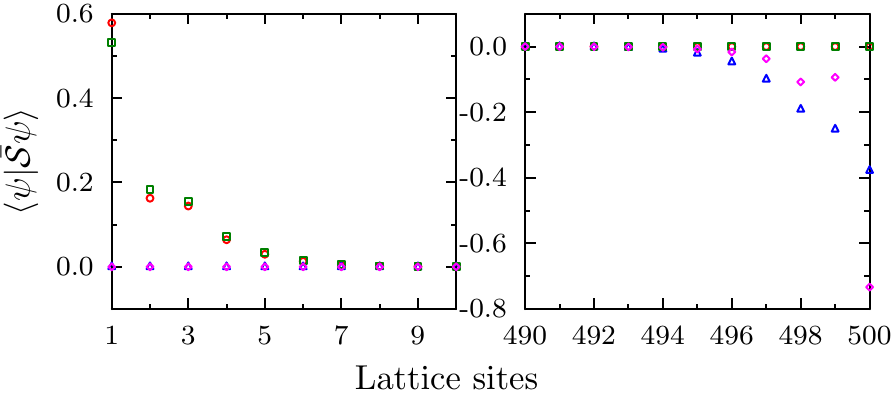}
\caption{(Color online) Local expectation value of the chiral operator $\bar{\mc S}$ in the zero-energy eigenstates at the first 10 and the last 10 sites of a $L=500 a$ wire. 
Two CBS, positive chiral eigenstates are localized at the right edge (red circles and green squares). The remaining two CBS, negative chiral eigenstates are localized at the left edge (blue triangles and magenta diamonds). The total ``chiral density'' obeys Eq.~(\ref{chiralNorm}).}
\label{fig:chiralEdges}
\end{figure}

Both kinds of edge modes are eigenstates of the chiral symmetry operator $\bar{\mc S}$. Two CBS, the positive eigenstates are localized at one edge, while the negative eigenstates, at the other edge. If $|\psi\ra$ is a zero-energy eigenstate, one can represent the spatial distribution for the chiral operator expectation value $\langle \psi|\bar{\mc S}\psi\rangle$. If $|\psi\ra$ is normalized to one, then the following equation holds: 
\begin{equation}\label{chiralNorm}
\sum_j\la\psi|j\ra\la j|\bar{\mc S}\psi\ra=\pm 1,
\end{equation}
depending whether $|\psi\ra$ is a positive or negative eigenstate. The sum is carried over all lattice sites.
This yields at once a representation for the localization of the modes, as well as a proof of them being chiral. The result is shown in Fig.~\ref{fig:chiralEdges}.
Note that the local weight of the expectation value between two eigenstates at two different edges is not correlated. There is a degree of freedom in distributing this weight for degenerate states, resting on the specifics of the diagonalization procedure. Nevertheless, the global Eq.~(\ref{chiralNorm}) always holds.

In conclusion, we have found a trivially topological phase with fragile midgap edge states which are eigenstates of the chiral operator. These states are removed from zero energy by chiral symmetry breaking perturbation. Also they can disappear by driving the system in a Majorana phase.

\subsection{Many-body ground-state degeneracy}
\label{subsec:many-body-gs-degeneracy}

Until now, we have only described the zero-energy edge modes from the perspective of single-particle physics, without touching on the question of the state filling. This subsection clarifies the way to count the ground states expected in the Creutz-Majorana model at the transition from fractionally charged solitons, to the CBS in the superconducting phase, and, subsequently, to Majorana bound states.

\begin{figure}[t]
\includegraphics[width=\columnwidth]{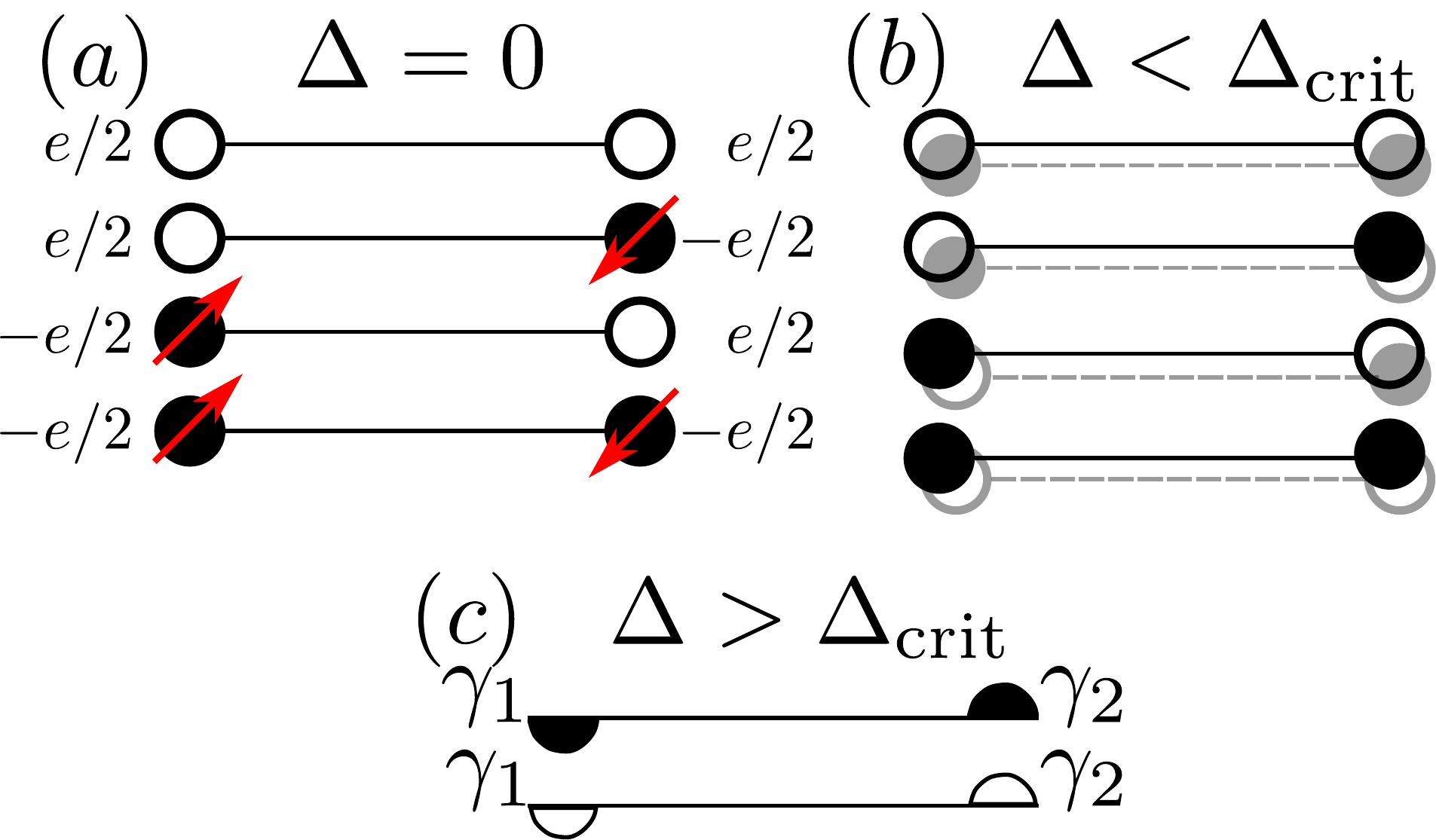}
\caption{(Color online) Different phases in the system under increasing pairing $\Delta$. The images stand for the possible ground states forming in the CM wires by studying the zero-energy states' occupation. The filled (empty) circles represent filled (empty) electronic states. (a) At zero pairing, i.e., in the Creutz model, there are fractionally charged bound states forming at the two ends of the wire. An electron filling a state at the left (right) edge has $\sigma_2$ projection spin $1/2$ ($-1/2$). (b) For finite pairing $\Delta$, but below the topological transition $\Delta<\Delta_{\rm crit}$, there are CBS zero modes. In the BdG there is a doubling of states (represented in gray) near zero energy. However, this redundant doubling does not modify the number of physical states, which remain four. The charge is no longer a good quantum number. (c) After crossing the transition to the Majorana phase ($\Delta>\Delta_{\rm crit}$), there are again just two states in the spectrum, two Majorana fermions forming at the two system edges. These states can be either filled or empty by with a \textit{single} electron. This is represented by having a fractionalized empty or filled electronic state at the two edges of the system.
}
\label{fig:occ}
\end{figure}

In the absence of superconductivity, the Creutz model provides nontrivial insulating phases with zero-energy states. These states can be either filled or empty with electrons. For long wires and at half filling, there will be no energy cost in having either of the states filled or empty. 
Therefore, while there are two states in the spectrum at zero energy, the ground state will be fourfold degenerate, with four possible configurations in the occupation of the zero modes (Fig.~\ref{fig:occ}).

The states at positive energy can be thought as states carrying positive charges. Due to the chiral symmetry, each filled negative energy state has an empty positive energy partner. Therefore, when there is a single electron occupying one of the two midgap states, the overall charge of the system is zero. When both zero-energy states are filled, there will be an excess of one electron charge $-e$ in the system. Because the charge of a midgap state can change only by one, it follows that each edge state will carry a fractional charge $-e/2$. Similarly, when both states are empty, the total energy does not change, and each edge state carries a charge $e/2$.\cite{Heeger1988}

Moreover, the states are eigenstates of the chiral operator $\sigma_2$. Here, the left modes  have spin projection up and the right states, spin projection down. Therefore, when both states are either filled or empty, the spin of the zero-energy pair is zero. When only the one state is filled, the total charge vanishes, but the ground state will have spin $1/2$.

Under the addition of superconductivity, the $U(1)$ charge symmetry is broken. However the system maintains its chiral symmetry and there will be CBS edge modes at zero energy.
In the BdG formalism, the number of states is doubled. Therefore, in the BdG spectra of the CM model, there will be four zero-energy end states.
Nevertheless, the degeneracy of the many-body ground state does not change. There are still four configurations, because the quasiparticle occupation numbers are not independent between particle-hole symmetric states.

While the states remain at zero energy, the ground-state is fourfold degenerate, either in the nontrivial insulating Creutz model, or in the CBS phases of the superconducting model. Breaking the chiral symmetry, allows  states to be moved away from zero energy, while remaining localized in bulk energy gaps, as it was shown in different nanowire systems.\cite{Gangadharaiah2011, Klinovaja2012a} Nevertheless, the ground state becomes in this case unique as for the trivially gapped system.

When crossing the topological transition to the Majorana phase, the BdG spectrum will reveal the existence of two Majorana solutions bound at zero energy at the ends of the wire. The ground state is also twofold degenerate. Two pair of states can be either filled or empty with a single electron. In contrast to the fractionally charged solitons, the electron itself can be viewed as divided between the edge states of the system in the form of two Majoranas. This is pictorially represented in Fig.~\ref{fig:occ}.

The upshot of this section is that the topological transition from CBS to Majorana fermions can be tracked either by looking at the spectrum of the BdG Hamiltonians, or from the many-body ground-state degeneracy.
These two view points must be clearly distinguished in the following, when interactions are treated either in mean-field theory, or by using DMRG techniques. While in mean field one has access to BdG Hamiltonians, and to their spectral properties, the DMRG obtains the many-body ground state.

\section{Creutz-Majorana-Hubbard Model}
\label{sec:CMH}
The Creutz-Majorana model on a finite-size chain hosts either zero-energy solitonic (CBS) states,  or Majorana modes, depending on the strength of the superconducting pairing $\Delta$. The aim of this section is to investigate the effects of repulsive onsite interactions on these edge modes and to obtain the topological phase diagram of the model. To that end, we consider the following Hamiltonian of the Creutz-Majorana-Hubbard (CMH) model: 
\begin{equation}
H=H_{\rm CM}+H_I.
\label{eq:cmh-model}
\end{equation}
The first term, $H_{\rm CM}$, represents the Hamiltonian of the Creutz-Majorana model from Eq.~(\ref{HCM}), and $H_I$ contains the Hubbard interaction between onsite electronic densities of opposite spin
\begin{equation}\label{Hint}
H_I=U\sum_jn_{j\up}n_{j\down},
\end{equation}
with $U$ being the interaction strength. 

\begin{figure}[t]
\includegraphics[width=\columnwidth]{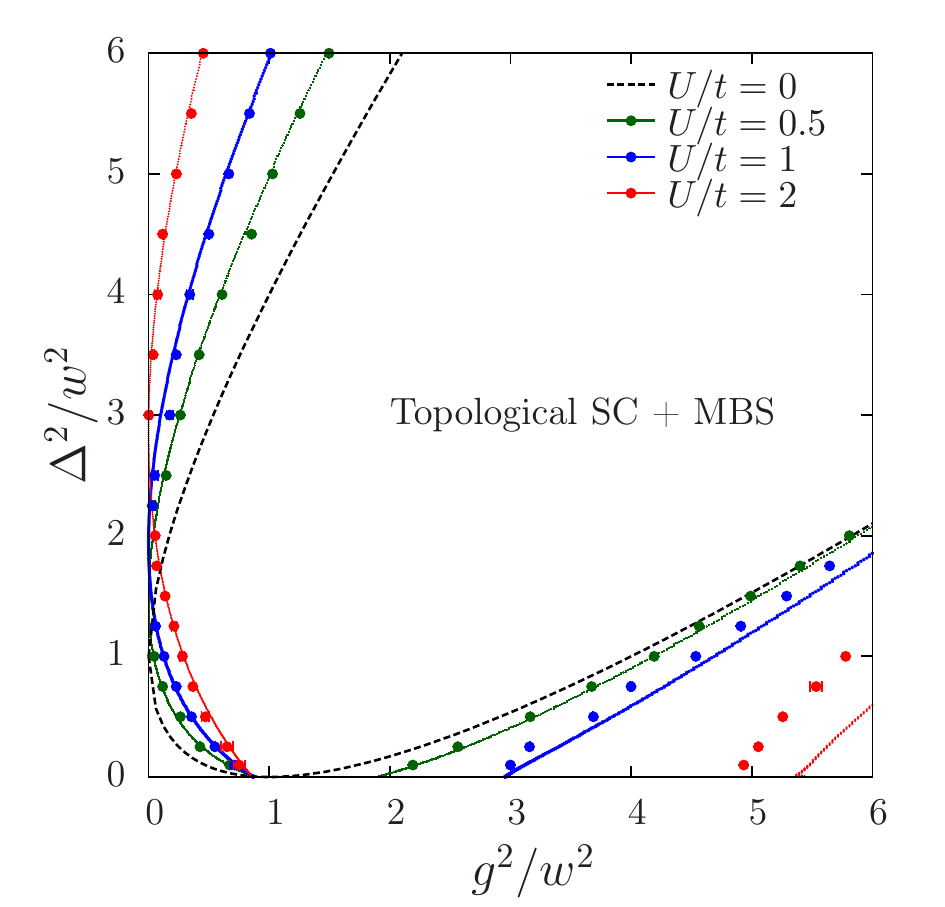}
\caption{(Color online) Topological phase diagram of the Creutz-Majorana-Hubbard model at different interaction strengths. The dotted line represents transitions between topological phases obtained at the mean-field level, while the dots represent critical points from the DMRG calculations. The topological superconductor phase which can host Majorana modes (MBS), represented in the central region, increases under the effects of interactions. The region at large ``spin-orbit'' $|g|$ and weak pairing $|\Delta|$ is a symmetry-protected topological phase. The rest of the outer regions are trivial superconductors with a unique ground state.
The agreement between mean field and DMRG breaks down at very large $g$, for strong interactions.}
\label{fig:MFDMRG}
\end{figure}

In this section, the CMH model is studied using a combination of self-consistent Hartree-Fock theory and extensive DMRG simulations. Let us begin by summarizing the main results and defer the details to the following subsections.

Figure~\ref{fig:MFDMRG} shows the topological phase diagram for different values of interaction $U$, combining results from the mean-field analysis and DMRG simulations. The agreement between both approaches is overall very good.
The main standout feature of the phase diagram is that the topological phase and its associated Majorana bound states at zero energy are robust to interactions.
In fact, the parameter range for which a topological Majorana phase is stabilized globally expands, upon increasing the Hubbard coupling. However, there is also a region in the phase diagram ($|\Delta/w|,|g/w|<1$), where the Hubbard interaction is detrimental to the Majoranas. Here their existence domain decreases with increasing interaction strength.
The mean-field phase diagram captures the topological transitions of the model rather accurately for small and moderate interactions. 
At strong interactions, the mean field keeps a very good estimate of the topological transition at small $g/w$, while at large $g/w$ it tends to overestimate the extension of the Majorana phase.
The ``spin-orbit'' coupling $g$ tends to delocalize the electrons and leads to quantum fluctuations in the particle number. This explains, at a qualitative level, the divergence of the mean field results from the ``exact'' DMRG results at large ``spin-orbit'' coupling.

\textit{A priori}, this overall very good agreement may seem surprising in a quantum 1D system, where fluctuations are expected to be pronounced.
However, let us recall that U(1) charge and the SU(2) spin-rotation symmetries  are  broken in this model, and hence the system acquires a certain stiffness to fluctuations, as observed in nanowire models  \cite{Stoudenmire2011} which also break these symmetries. 

\begin{figure}[t]
\includegraphics[width=\columnwidth]{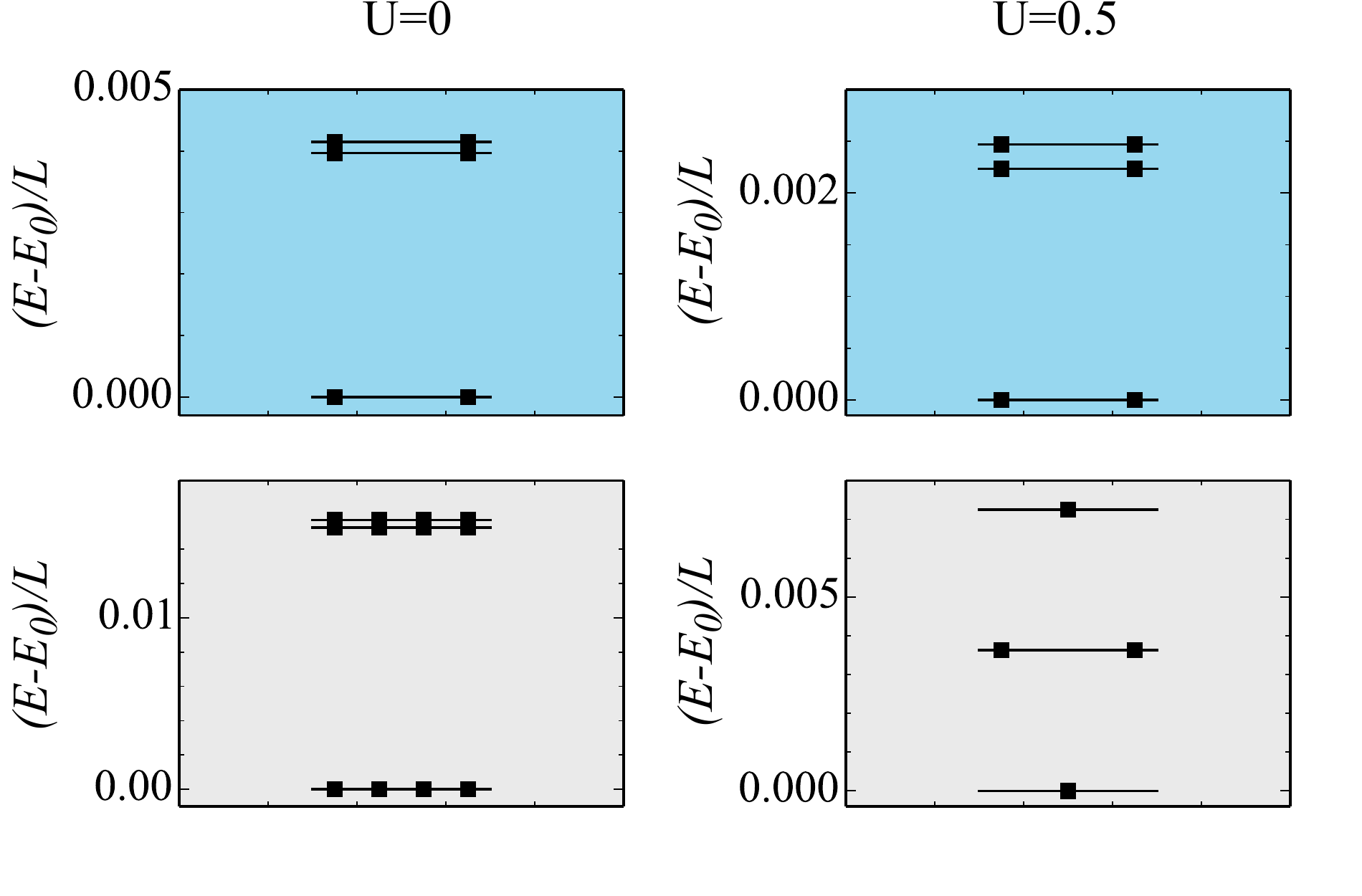}
\caption{(Color online) DMRG results for the energy spectrum measured from the ground-state energy $E_0$ for $L=64a$ open wires. Top row: $\Delta^2=0.0625$, $g^2=1$.  The double degeneracy in the topological phase with MBS survives the addition of interactions $U$. 
Bottom row: $\Delta^2=0.0625$, $g^2=20$.  The fourfold energy degeneracy of the phase hosting CBS is removed by interactions $U$.}
\label{fig:dmrg_spectrum}
\end{figure}

Crucially, the interactions have a different effect on the two types of zero-energy states localized at the edges of the wire. 
The chiral bound states from the large $g/w$ phase (cf. Fig.~\ref{fig:phdiag}) are removed from zero energy. Recall that for $U=0$, there is a fourfold zero-energy degeneracy in the BdG energy spectrum as well as a fourfold degenerate many-body ground state.  
At the mean-field level, upon increasing the interaction strength, the zero-energy CBS acquire a finite energy. Further increase of the interaction leads to a decrease of the bulk gap and pushes the end states into the bulk continuum. (Fig.~\ref{fig:zeroES}). As soon as the CBS aquire a finite energy, the many-body ground-state degeneracy is lifted. DMRG results confirm that the ground state becomes unique, corresponding in the BdG picture to having all negative quasiparticle states filled (Fig.~\ref{fig:dmrg_spectrum}). 
Therefore, the degeneracy of the ground state in the CBS phase is not protected with respect to Hubbard interactions. However, the system remains in a ``nontrivial'' phase which is protected by a product of fermion parity and inversion symmetry. This phase is characterized by a bulk topological invariant. 

In contrast, the Majorana end states generally remain pinned at zero energy as expected from the phase diagram. Consequently, the ground state of the system remains twofold degenerate. As already remarked, only in a limited region with a small superconducting gap ($|\Delta/w|<1$) and small ``spin-orbit'' coupling ($|g/w|<1$) can Majorana fermions be destroyed by interactions (Fig.~\ref{fig:MFDMRG}).

\subsection{Mean-field study}
\label{subsec:MF}

At the mean-field level, the four-operator interaction $H_I$ is approximated by a  sum of two-operator interactions. In the present case, we choose the following Hartree-Fock decoupling of density-density interaction term from Eq.~(\ref{Hint}):
\begin{eqnarray}\label{decoup}
n_\up n_\down&\approx&
 n_\up\avg{n_\down}+n_\down\avg{n_\up}
+\delta c^\dag_\up c^\dag_\down+\delta^* c_\down c_\up\\
&&-c^\dag_\up c_\down\eta^*-c^\dag_\down c_\up\eta
-\avg{n_\up}\avg{ n_\down}-|\delta|^2+|\eta|^2\notag,
\end{eqnarray}
where the site index was neglected, since the decoupling involves only local terms at a given site.
The brackets $\langle\dots\rangle$ denote the expectation value in the ground state.
The site-independent parameters $\delta$ and $\eta$ are defined:
\begin{equation}
\delta=\la c_\down c_\up\ra\quad\eta=\la c^\dag_\up c_\down\ra.
\end{equation}
A finite anomalous pairing $\delta$ indicates intrinsic superconducting correlation produced by repulsive interactions $U$. Note that the Hamiltonian $H$, already has an induced (extrinsic) superconducting pairing $\Delta$, which is independent on the interactions $U$. The finite $\eta$ represents the tendency towards a polarization in the $x$ direction.

Let us also introduce the onsite electronic density $\rho$ and the onsite magnetization $m$ (in the $z$ direction),
\begin{equation}
\rho=\la n_\up+n_\down\ra,\quad m=\la n_\up-n_\down\ra,
\end{equation}
where again these quantities are uniform.
Consequently, the ground state of the system is characterized by the expectation values $(\rho,m,\eta,\delta)$.
The  Wick theorem is verified by evaluating both sides of Eq.~(\ref{decoup}) in the ground state.

Because the specific decoupling in Eq.~(\ref{decoup}) is site independent, the implicit assumption that we have made is that the ground state can develop only ferromagnetic instabilities. However, the DMRG simulations do not reveal evidence of antiferromagnetic instabilities, so in this approximation we neglect antiferromagnetic order parameters or more complicated nonlocal decoupling terms. A further simplification in the present mean-field theory is to consider only real order parameters.

Let us consider the mean-field Hamiltonian as describing a periodic chain, where the ground state is unique.
Therefore, a Fourier transform allows us to determine the momentum-space mean-field Hamiltonian $\mc H(k)$:
\begin{equation}
H_{\rm MF}=\frac{1}{2}\sum_k C^\dag_k\mc H(k)C_k+E_I.
\end{equation}
The interaction energy $E_I$ reads as
\begin{equation}\label{intE}
E_I=LU\bigg(
\frac{\rho}{2}-\frac{\rho^2}{4}+\frac{m^2}{4}+\eta^2-\delta^2
\bigg),
\end{equation}
and the BdG Hamiltonian $\mc H(k)$,
\begin{eqnarray}\label{meanFieldK}
\mc H(k)&=&t\sin k\sigma_3\tau_0+(w-U\eta-g\cos k)\sigma_1\tau_3\notag\\
&&-(\Delta+U\delta)\sigma_2\tau_2+\frac{U\rho}{2}\sigma_0\tau_3-\frac{Um}{2}\sigma_3\tau_3,
\end{eqnarray}
where it becomes clear that interactions renormalize the onsite terms. The exchange term $\eta$ modifies the $x$ polarization $w$ and the anomalous pairing $\delta$ renormalizes the proximity-induced pairing $\Delta$. In addition, the interaction also introduces a chemical potential $U\rho$ and a $z$-polarization term $Um$.
The $m$ term would act as a Zeeman term, breaking the time-reversal symmetry. 

The mean-field Hamiltonian~(\ref{meanFieldK}) is quadratic and can be diagonalized in a basis of quasiparticles as: 
\begin{equation}
H_{\rm MF}=\frac{1}{2}\sum_k\Gamma^\dag_k\mc H_{qp}\Gamma_k+E_I,
\end{equation}
with $\Gamma^\dag_k
=(\gamma^\dag_{ka},\gamma^\dag_{kb},\gamma_{-ka},\gamma_{-kb})$, and the Hamiltonian matrix, $\mc H_{qp}={\rm diag}(\e_a,\e_b,-\e_a,-\e_b)$. The quasiparticle operators $\gamma^\dag_{a,b}$ create quasiparticle excitations at positive energies $\e_{a,b}>0$. Also, note that the spectrum exhibits the particle-hole symmetry with a spectrum symmetric about the zero energy. 

The system ground state $|\text{GS}\ra$ is annihilated by the quasiparticle excitations. Therefore, it can be represented as
\begin{equation}\label{GS}
|\text{GS}\ra=\prod_k\gamma_{ka}\gamma_{kb}|0\ra,
\end{equation}
where $|0\ra$ is the vacuum of conventional fermionic annihilation and creation operators. The usual creation and annihilation operators can be decomposed on the basis of the quasiparticle excitations $\Gamma$ with coefficients given by the eigenvectors of the Hamiltonian $\mc H(k)$.

The order parameters for the mean-field Hamiltonian are found using a self-consistent procedure. At the start, the order parameters are assumed to be zero. After diagonalizing the mean-field Hamiltonian, the order parameters are evaluated in the ground state $|\text{GS}\ra$ given by Eq.~(\ref{GS}). This allows us to determine the interaction energy $E_I$, and, consequently, the total Hartree-Fock energy, given by the occupied quasiparticle energy levels, plus the interaction energy ($E_{\rm HF}=E_I-\e_{a}-\e_{b}$). The resulting order parameters are fed back into the mean-field Hamiltonian and a new ground state is obtained. Therefore, new order parameters can be calculated. The cycle of operations is repeated until self-consistency is reached (i.e., the ground-state energy settles to a minimum value among cycles). As a consequence, this procedure generates a final set of order parameters $(\rho,m,\eta,\delta)$ which completely characterize the mean-field Hamiltonian.

\begin{figure}[t]
\includegraphics[width=0.75\columnwidth]{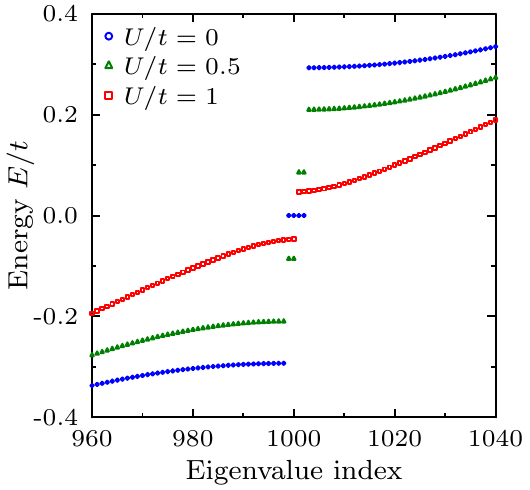}
\caption{(Color online) Energy eigenvalues of a finite Creutz-Majorana model at three different interaction strengths. The interaction pushes the CBS located in the trivial superconducting region at finite energy, rendering the ground state unique. Model parameters are $\Delta^2/w^2=0.5$, $g^2/w^2=4$ and the wire length in $ L=500a$.}
\label{fig:zeroES}
\end{figure}

In the mean-field calculations, we consider wires of lengths $L=100a$ at several interaction strengths. The self-consistent procedure is performed over a $500\times 500$ point grid in the $(\Delta/w,g/w)$ parameter space to obtain the order parameters of the mean-field Hamiltonian.

Crucially, the self-consistent analysis (backed by DMRG results) reveals that there is no polarization developing in the $z$ direction. The vanishing Zeeman term $Um$ implies that the mean-field Hamiltonian is $\mc T$ invariant. Consequently, it can still be classified in the BDI class, as a noninteracting and translation-invariant Hamiltonian, which can be treated similarly to the Creutz-Majorana model in Sec.~\ref{sec:CM}.
Therefore, information about its topological phases can be extracted from the topological invariant:
\begin{equation}
\mc M=\sgn\big\{\big[\bar w^2+g^2-\bar\Delta^2-U^2\langle n_\up\rangle \langle n_\down\rangle\big]^2-4\bar w^2g^2
\big\},
\end{equation}
where the renormalized on-site terms read as
\begin{equation}
\bar w = w-U\eta,\quad\bar\Delta=\Delta+U\delta,\quad\langle n_\up\rangle \langle n_\down\rangle=\frac{\rho^2}{4}.
\end{equation}
The topological phase diagram from Fig.~\ref{fig:MFDMRG} follows by plotting the transition lines where the Majorana number $\mc M$ changes sign.

The properties of the edge states due to interactions can be explored at mean-field level. Retaining the interaction-renormalized order parameters obtained for the periodic system, one can study wire models with open edges. Hence, the existence of MBS and CBS can be tested in comparison with further DMRG results (Sec.~\ref{subsec:spectral properties}).

At the mean-field level, interactions produce a finite uniform scalar potential $U\rho\sigma_0\tau_3$, which breaks the chiral symmetry $\bar{\mc S}$. Therefore, the zero-energy CBS inherited from the Creutz model are no longer protected. As shown in Sec.~\ref{subsec:CtoM}, this leads to a removal of the zero-energy degeneracy for CBS states. Fig.~\ref{fig:zeroES} shows the lifting of the CBS modes from zero energy under increasing interaction strength $U$.

\subsection{DMRG Study}
\label{subsec:dmrg}
In this section,  the DMRG study of the ground-state phases and phase transitions of the Creutz-Majorana-Hubbard model is discussed.
We first employ the DMRG  method~\cite{White1992} in an MPS formulation~\cite{Schollwock2011} to obtain the three lowest-lying levels of the energy spectrum for open chains of length $L=64a$.
This is done successively, by first obtaining the ground state in MPS form, and then performing a new DMRG calculation optimizing an MPS which is orthogonal to the previous ones.\cite{Schollwock2011}
The resulting spectra, shown in Fig.~\ref{fig:dmrg_spectrum}, illustrate the fate of the MBS and CBS when interactions are considered.
The results are consistent with mean-field theory of the open chain. When starting from a phase with CBS modes, interactions remove the fourfold degeneracy  and yield a unique ground state (bottom-right panel in Fig.~\ref{fig:dmrg_spectrum}). It corresponds to having all negative quasiparticle states occupied. In contrast, the twofold degeneracy of the ground state in the Majorana phase is robust to interactions.

The above procedure on open chains is, however, rather cumbersome, especially when  close to the critical points where the correlation length rises and finite-size effects from the coupling of edge modes can become relevant.
For this reason, we employ the more efficient iDMRG algorithm~\cite{Mcculloch2008,Schollwock2011,Kjall2013} to obtain an MPS representation of the ground-state wave function for an {\it infinite} system
(see Ref.~\onlinecite{Kjall2013} for the details of the algorithm).
We conserve the fermion number parity in our simulations, and use a maximum bond dimension up to $\chi\approx500$ when calculating ground states.
Obtaining the energy spectrum of an infinite chain is problematic. However, the entanglement spectrum is readily available and also contains information about the degeneracies of the system.\cite{Li2008,Pollmann2010,Turner2011,Fidkowski2011}

An  MPS representation of a wave function $|\psi\rangle$ can be cut anywhere in the chain~(including across a physical site), yielding the Schmidt decomposition

\begin{align}
| \Psi \rangle = \sum_\alpha \Lambda_\alpha |\alpha\rangle_L \otimes |\alpha \rangle_R,
\end{align}
where $|\alpha\rangle_{L/R}$ represent the states to the left/right respectively of the cut.
Taking $|\alpha\rangle_{R}$, these are the eigenstates of the reduced density matrix for the right
half of the system, $\rho_R={\sf Tr}_L(|\psi\rangle\langle\psi|)$. 
The \textit{bulk} entanglement spectrum,
\begin{align}
\{ \epsilon_\alpha \}= -2\log(\Lambda_\alpha),
\end{align}
reflects how the different Schmidt states contribute to the wave function~$|\psi\rangle$. 
It can be interpreted as the spectrum of the so-called entanglement Hamiltonian $\mathcal{H}_E=-\log\rho_R$ which contains information about ``artificial edges'' of semi-infinite chains.
The symmetries that protect the topological degeneracies in the energy spectrum associated with open edges also protect degeneracies in the entanglement spectrum.\cite{Pollmann2010,Turner2011,Fidkowski2011} In the following, we work in units where $w=t=1$.

\begin{figure}[t]
\centering
\includegraphics[width=\columnwidth]{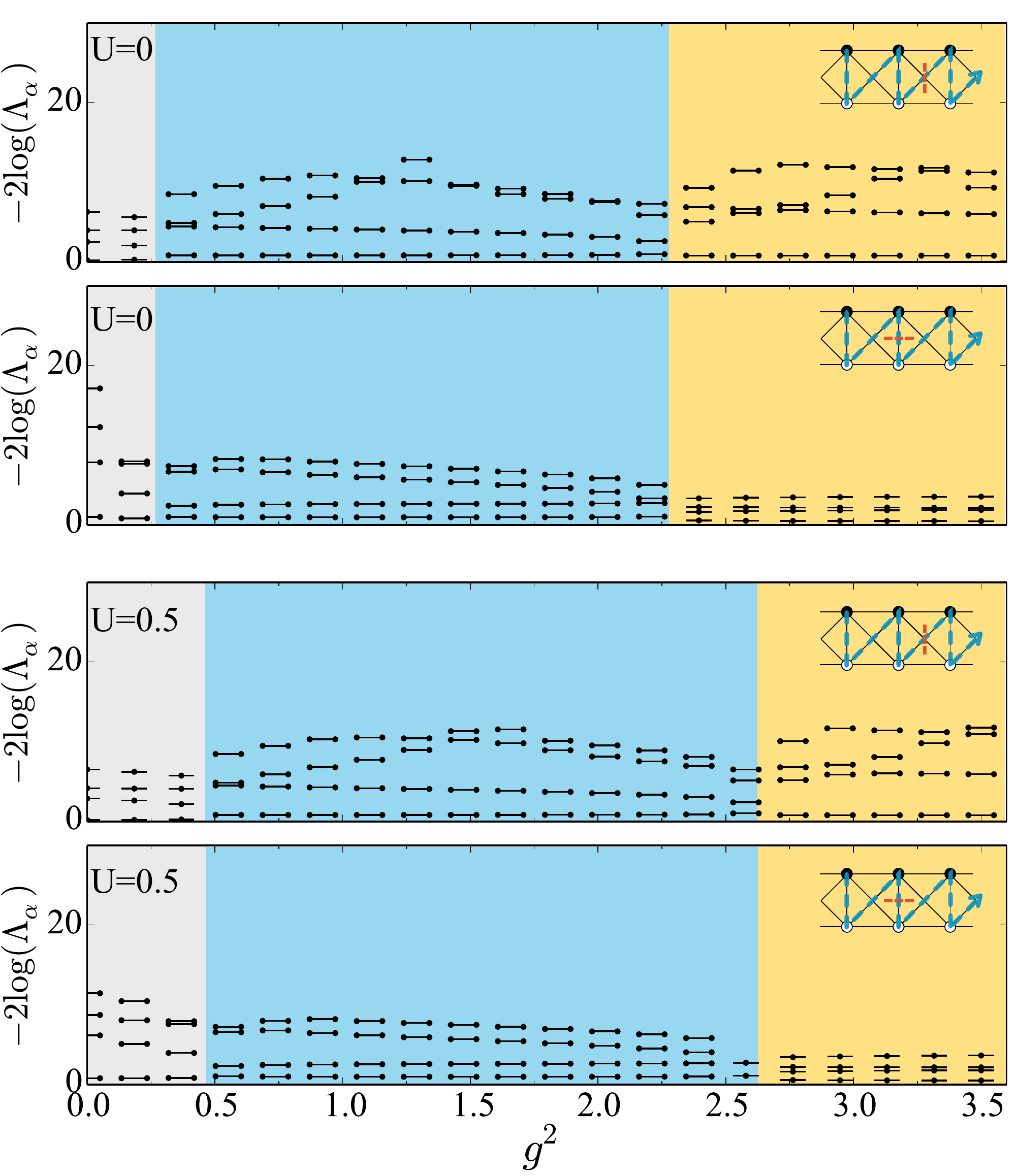} 
\caption{
(Color online)
Phases found in the Creutz-Majorana-Hubbard model at fixed $\Delta^2=0.25$, from  iDMRG results. 
By increasing $g$ we sweep through the trivial  phase, the TSC and an inversion symmetry-protected topological (SPT) phase.
The bulk entanglement spectrum is obtained from two different cuts of the chain, shown in the diagrams in the right corners.
The topological phase with Majorana edge states (blue) displays a twofold entanglement degeneracy for both kinds of cuts, unlike the phases surrounding it.
}
\label{fig:sweep}
\end{figure}

The different phases found in Fig.~\ref{fig:sweep} for $\Delta^2=0.25$ can be distinguished by their entanglement spectrum, whose features survive the addition of interactions.
We perform two different cuts of the chain, as illustrated in the insets of Fig.~\ref{fig:sweep} : one of the cuts corresponds to separating two physical sites; and the other can be seen as the act of polarizing a single site.

The topological phase with Majorana end states (blue), is distinguished from the other phases by its twofold entanglement degeneracy for both kinds of cuts.
This degeneracy is due to the fractionalization of the fermion parity symmetry $\mathcal{Q}=e^{i\pi \sum_{j}n_{j}}\label{Q}$ and can not be lifted unless the system undergoes a phase transition, as discussed in detail in Ref.~\onlinecite{Turner2011}.

The trivial phase at small $g$ is only degenerate for cuts inside a physical site.
The phase at large $g$ (yellow) displays entanglement degeneracy only for a vertical cut, separating two physical sites.
This degeneracy is protected by the product of the fermion parity and inversion symmetry, i.e., by  $\mathcal{\tilde{I}}=\mathcal{Q}.\mathcal{I}$ with $\mathcal{I}$ being the inversion of the chain at a bond.
In order to distinguish this phase from the trivial phase at small $g$, we employ a method similar to the one used to classify topological phases in spin chains.\cite{Pollmann2012}
The basic idea is to find the cohomology class to which the ``fractionalized" representation $U_{\Sigma}$ of a symmetry $\Sigma$ belongs.
The matrices $U_{\Sigma}$ are representations of the symmetry $\Sigma$ acting on the Schmidt states $|\alpha\rangle$.
In our case, assuming a two-site unit  cell, we use the symmetry $\mathcal{\tilde{I}}$  to calculate $U_{\mathcal{\hat{I}}}$. 
We find that the topological invariant $U^{\phantom*}_\mathcal{\tilde{I}}U^{*}_\mathcal{\tilde{I}}=-1$ and thus the phase is a topological phase, while in the trivial phase for small $g$ we find  $U^{\phantom*}_\mathcal{\tilde{I}}U^{*}_\mathcal{\tilde{I}}=+1$.
A direct consequence of $U^{\phantom*}_\mathcal{\tilde{I}}U^{*}_\mathcal{\tilde{I}}=-1$ is the observed even degeneracy in the entanglement spectrum  (see Ref.~\onlinecite{Pollmann2012} for details).  
The spatial inversion symmetry is broken at the edges of an open chain and thus there will be no signatures of this phase in the energy spectrum of Fig.~\ref{fig:dmrg_spectrum}.
The existence of the  CBS end states at $U=0$ is stabilized by the presence of further symmetries, which are broken by interactions (as discussed above).

\begin{figure}[t]
\centering
\includegraphics[width=\columnwidth]{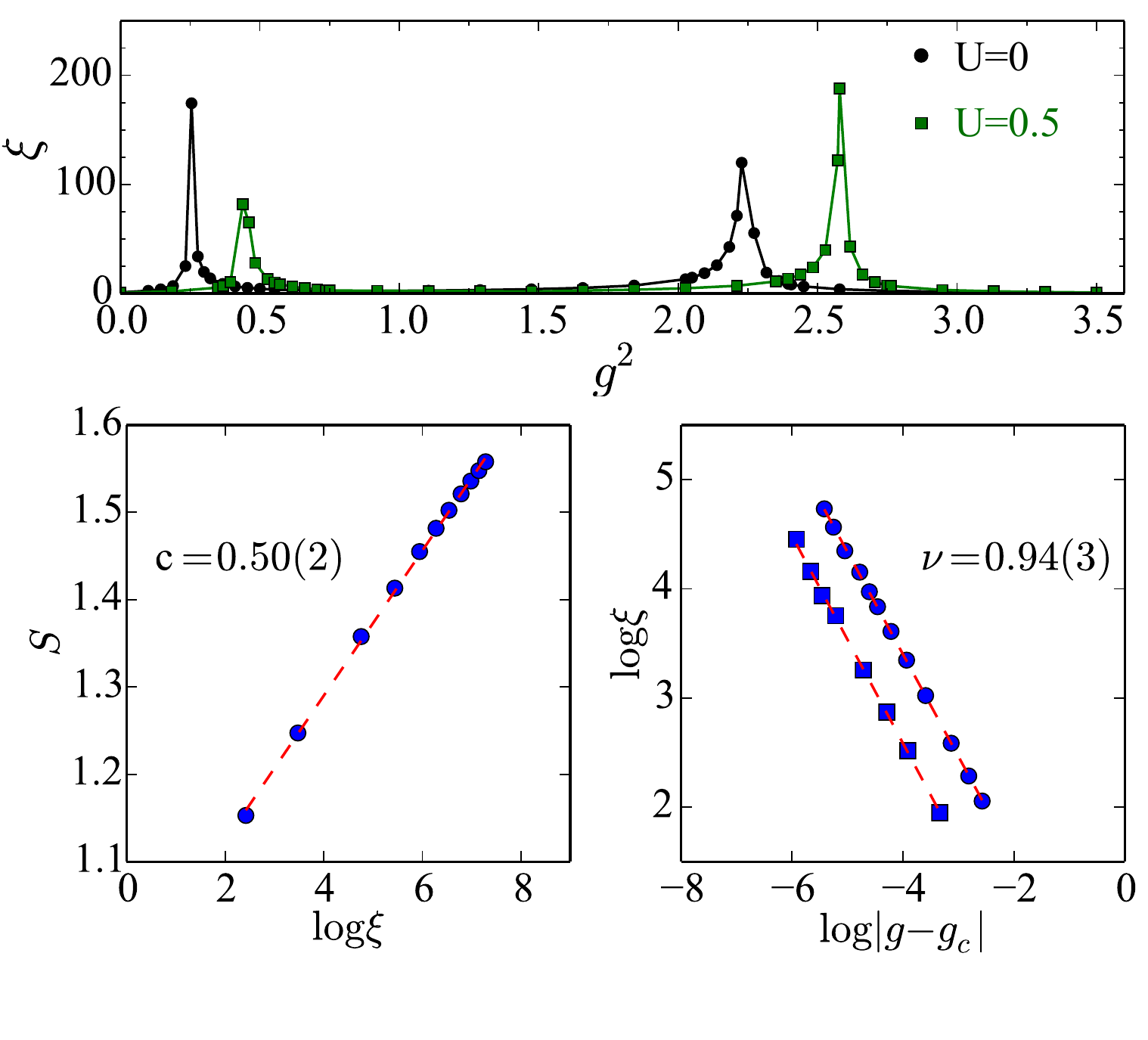} 
\caption{
(Color online)
Top: Topological transitions are tracked by peaks in the correlation length $\xi$ at different interaction strengths.
Bottom: Finite-entanglement scaling of iDMRG results of the phase transition at $U$=$1$, $\Delta^2=0.25$ and $g_c=1.80840(4)$.
Left: Scaling  of the entanglement  entropy with the correlation length at $g_c$. 
Right: Scaling of the correlation length with the distance to the critical coupling $g_c$.
Results are representative of all transitions observed and consistent with the Ising universality class. }
\label{fig:critical-scaling}
\end{figure}

The boundaries between different phases correspond to gapless points in the thermodynamic limit.
We extract the correlation length $\xi$ of the system from the spectrum of the transfer matrix of the MPS. 
A sharp increase near the boundaries between phases is observed in Fig.~\ref{fig:critical-scaling} (top), both with and without interactions.
This reflects the diverging behavior of the correlation length close to the critical point, cutoff by the finite bond dimension used, as discussed in Ref.~[\onlinecite{Kjall2013}].
We therefore take the location of the $\xi$ maxima as an estimate for the position of the critical points.

The transition between the TSC phase and the topological phase stabilized by $\hat{\mathcal{I}}$  at  $U=1$, $\Delta^2=0.25$ and $g^2\approx3.25$  in Fig.~\ref{fig:critical-scaling}(bottom row) is a representative case of \emph{all} the phase transitions  reported here.
In order to analyze the phase transitions, we employ a finite-entanglement scaling approach~\cite{Pirvu2012,Kjall2013} of the entanglement entropy $S$ and correlation length $\xi$. The entanglement entropy, $S=-\sum_\alpha\Lambda_\alpha^2\ln\Lambda_\alpha^2$,  also displays a divergent  behavior when approaching the critical point (not shown).
The spin-orbit term $g$ is first fine-tuned in order  to give the largest possible $\xi$ for $\chi\approx500$. This provides an estimate of $g_c$ and ensures that the system is in the finite-entanglement scaling region.
At this point, the entanglement entropy scales with the calculated correlation length, setting an effective length scale\cite{Calabrese2004},
\begin{align}
S=\frac{c}{6}\log(\xi) + s_0,
\end{align}
giving access to the central charge $c$ of the critical point.
A standard scaling form for the correlation length is also used,
\begin{align}
\log|g-g_c| = \nu\log\xi + b.
\end{align}
The extracted central charge $c$ and correlation-length exponent $\nu$ are in very good agreement with those of the 2D Ising universality class, $c=1/2$ and $\nu=1$.
All the transitions reported between the different phases, with and without interactions, belong to the Ising universality class.

\begin{figure}[t]
\centering
\includegraphics[width=\columnwidth]{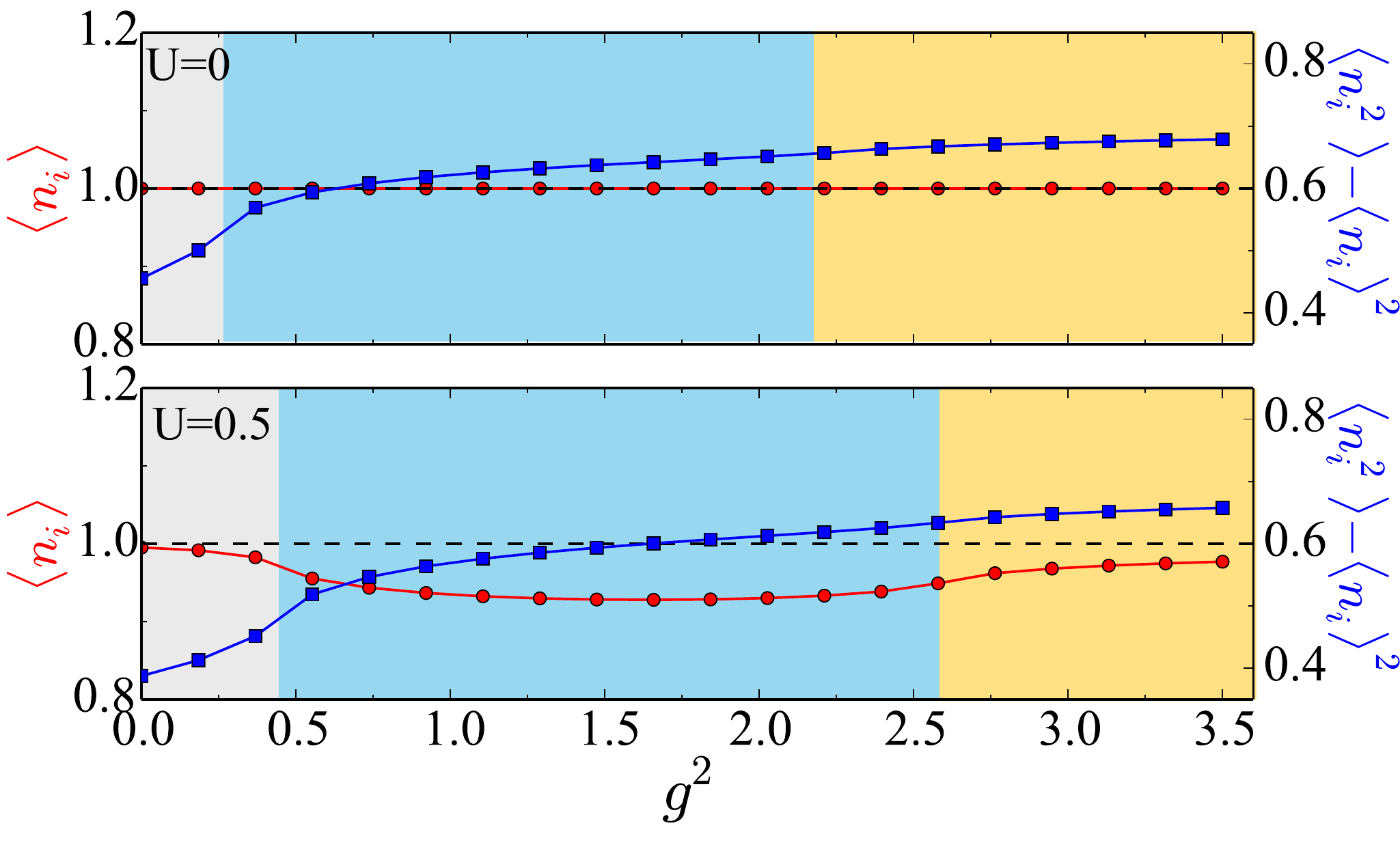} 
\caption{
(Color online)
Phases in the Creutz-Majorana-Hubbard model at $\Delta^2=0.25$ from iDMRG simulations.
Top:  $U$=$0$, The transition into the TSC phase is marked by a rise in particle-number fluctuations for simulations initialized at half filling $\langle n_i\rangle=1$.
Bottom: $U$=$0.5$, Repulsive interactions lower the average number of fermions in the system, thereby inducing an effective chemical potential, which lift the chiral bound states at large $g$.
}
\label{fig:sweep-fluctuations}
\end{figure}

In order to connect with the mean-field results of Sec.~\ref{subsec:MF}, we analyze
in  Fig.~\ref{fig:sweep-fluctuations} the  average fermion number per site and associated fluctuations.
For $U=0.5$ and  $g>0$ (bottom), the  fermion number deviates from half filling in order to reduce the Hubbard energy for on-site repulsions.
In the mean-field description, this corresponds to an effective (uniform) chemical potential, that lifts the chiral symmetry $\bar{\mathcal{S}}$ which protected the zero-energy chiral bound states.
Additionally, the increasing fluctuations of the fermion number with $g$ imply that the mean-field approach of Section~\ref{subsec:MF} becomes less accurate for large $g$.

\begin{figure}[t]
\centering
\includegraphics[width=\columnwidth]{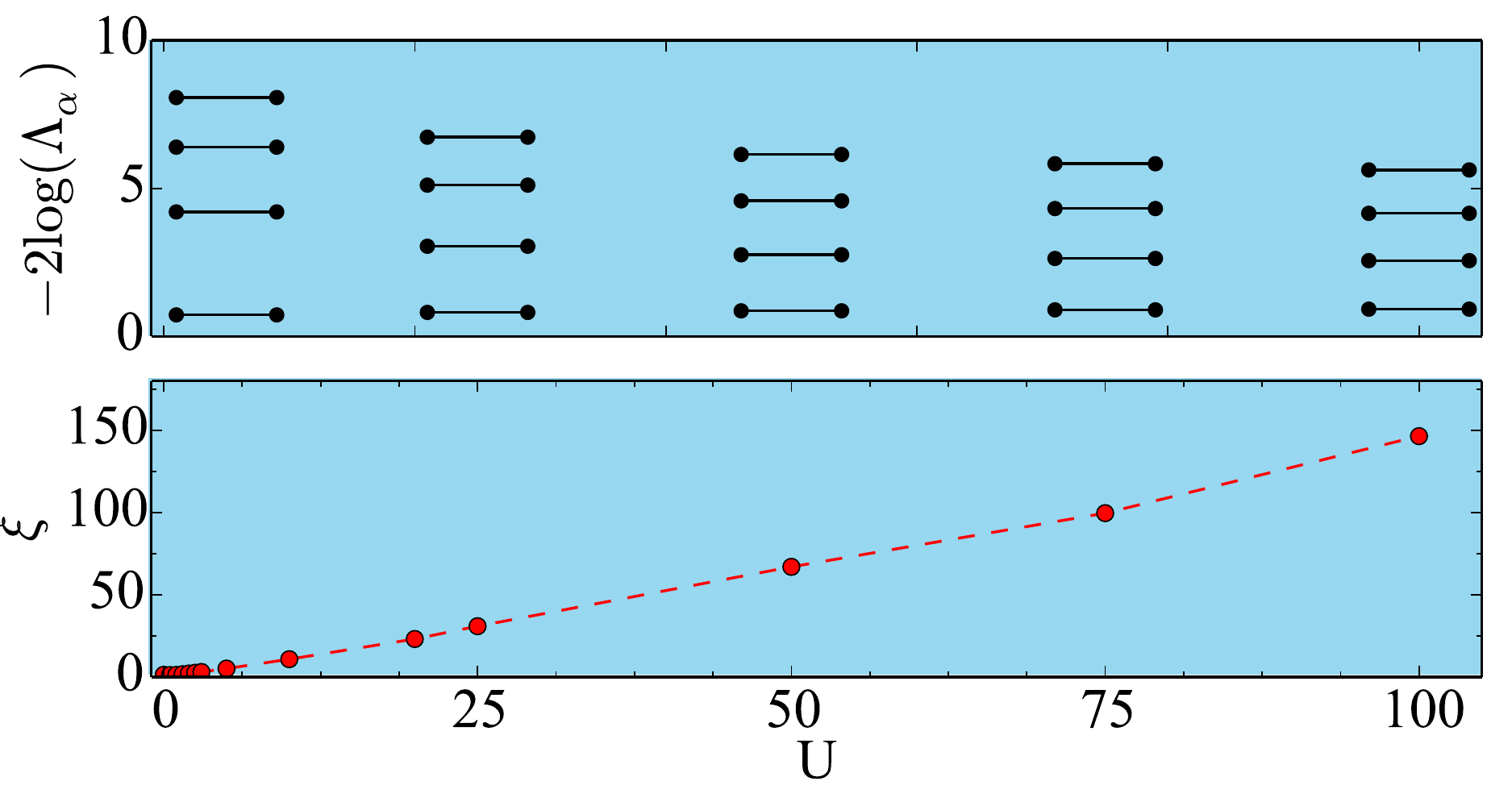} 
\caption{(Color online) Robustness of the topological phase to large interactions from iDMRG results. $g^2=\Delta^2=1$.
Top: The entanglement spectrum remains twofold degenerate.
Bottom: The correlation length increases but remains finite.}
\label{fig:large-U}
\end{figure}

The topological phase with MBS is extremely robust to strong interactions,  as shown 
by the persistence of the twofold degeneracy of the entanglement spectrum
in  Fig~\ref{fig:large-U}(top)  for $g^2=1$, $\Delta^2=1$ up to $U=100$.
The correlation length increases with increasing $U$ (bottom panel), but it is always found to converge to a finite value.
Hence, the bulk energy gap remains finite, albeit reduced by increasing interactions. 
These conclusions are in good agreement with those reached by Stoudenmire \textit{et al.}~\cite{Stoudenmire2011} for the effects of Hubbard interactions on a topological phase hosted by a nanowire model.
We have additionally checked that this phase is robust to a moderate chemical potential randomly distributed across a finite chain (not shown).


\subsection{Spectral properties}
\label{subsec:spectral properties}

The excitation properties of topological phases provide important clues to their nature, as well as insight into quantities which can be used experimentally to measure them.~\cite{Alicea2012,Beenakker2013}
For example, the spectral functions $A(k,\omega)$ can be used to track the closing of the bulk energy gap, which may signal a topological transition. 
Moreover, the \textit{local} spectral function $A(\omega)$  describes the local single-particle density of states at the edges of an open chain. This information could be used to detect Majorana fermions via a $2e^2/h$ quantized zero-bias signal in the differential conductance.\cite{Fu2009,Akhmerov2009,Law2009,Flensberg2010}
In this section, the spectral functions are calculated within mean-field theory and using MPS techniques.

\subsubsection{Excitation spectrum}
\begin{figure}[t]
\centering
\includegraphics[width=\columnwidth]{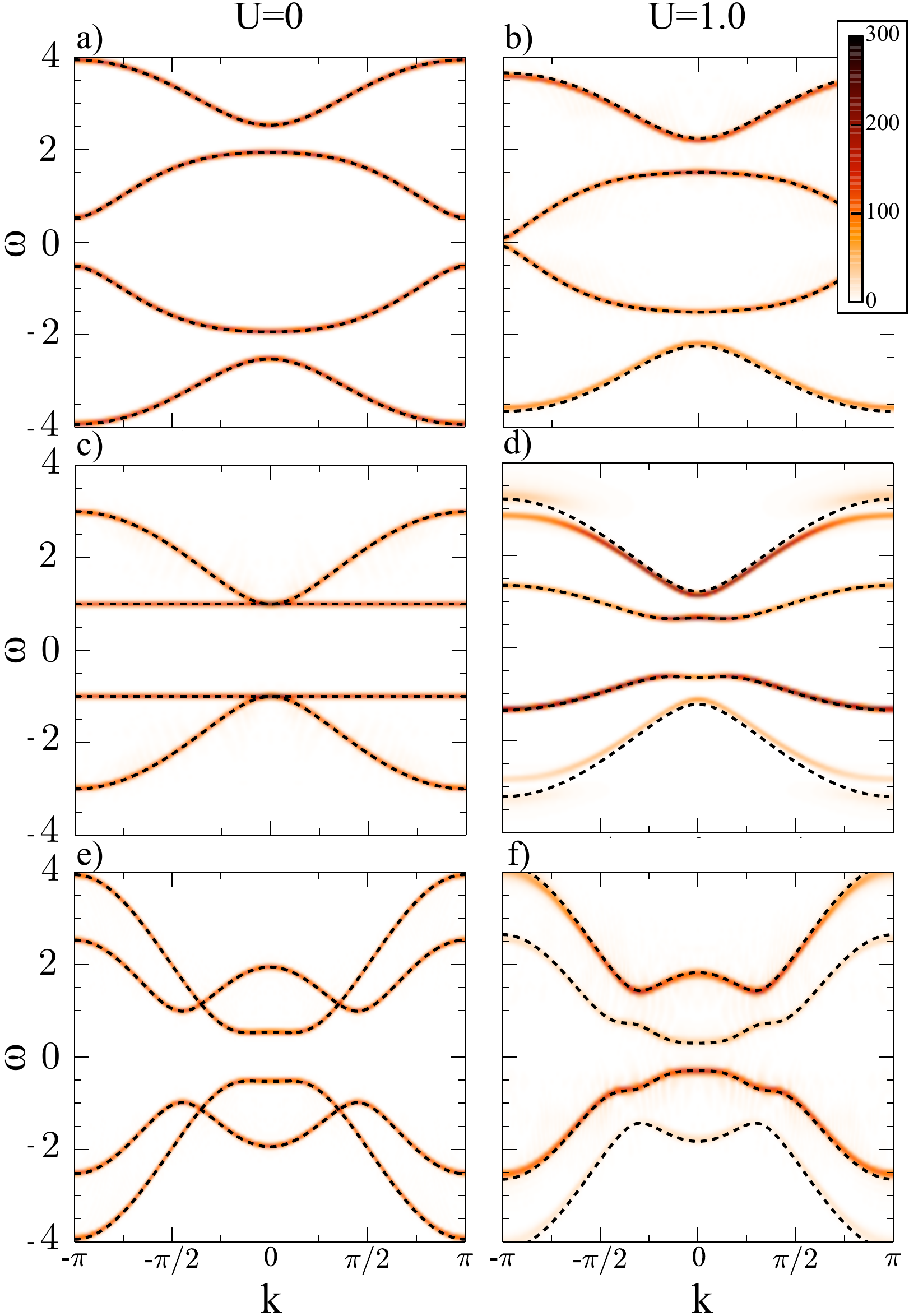} 
\caption{
(Color online)
Bulk spectral function $A(k,\omega)$ for $U=0$  (left column) and $U=1$ (right column). 
(a, b) Trivial phase for $g^2=0.5$, $\Delta^2=5$. 
(c, d) TSC  phase with Majorana edge modes for $g^2=\Delta^2=1$.
(e) Symmetry-protected topological phase with CBS edge modes and (f) without them for  $g^2=5,\Delta^2=0.5$. 
Color map is a TEBD calculation and dashed lines come from the mean-field theory.
}
\label{fig:akw}
\end{figure}

At the mean-field level, the excitation spectrum is obtained by solving for the eigenenergies of the BdG Hamiltonian from Eq.~(\ref{meanFieldK}) on a periodic chain. The self-consistent Hartree-Fock calculation results are represented in comparison with the DMRG results in Fig.~\ref{fig:akw}.

Spectral functions can be obtained with an MPS approach via the evolution of correlations in real time with the time-evolving block decimation (TEBD) method.\cite{Vidal2004,Vidal2007,White2004,White2008,Schollwock2011}
A fermionic operator $c$ or $c^\dagger$ (the spin index is suppressed for clarity) is applied at site $i_0$ to the ground state MPS of a finite chain.
The resulting excited MPS is evolved in real time, such that the hole and particle Green's functions, defined as
\begin{align}
G^{\sf h}=-i\langle  c^{\dagger}_ {i}(0) c^{\phantom \dagger}_{i_0}(t) \rangle,
\label{eq:GFs}
\\
G^{\sf p}=-i\langle  c^{\phantom \dagger}_{ i}(t) c^{\dagger}_{ i_0}(0) \rangle,
\label{eq:GFs-2}
\end{align}
are calculated separately.
A Fourier transform to momentum and frequency space gives the retarded single-particle Green's function\cite{Mahan2000} $G_{\sf ret}(k,\omega)$=$G^{\sf h}(k,\omega)+G^{\sf p}(k,\omega)$. The momentum-resolved spectral function is given by
\begin{align}
A(k,\omega)= -\frac{1}{\pi} \textrm{Im} [G_{\sf ret}(k,\omega)],
\label{eq:Akw}
\end{align}
which probes the energy spectrum associated with single-particle excitations in the bulk.
We let the matrix bond dimensions grow with time such that the truncation error is at most $10^{-5}$ per step. 
A fourth-order Suzuki-Trotter decomposition with $dt=0.04$ is employed, where the time scale is set by the inverse hopping $1/t=1$.
The maximum time is chosen such that the wave-front of the ``light cone" of correlations does not reach the edges of the system, resulting in a cutoff at small frequencies.
The sampled time is extended using linear prediction\cite{White2008} to improve the resolution in the limit: $\omega\rightarrow 0$.

The bulk spectral functions $A(k,\omega)$ for the different phases found are shown in Fig.~\ref{fig:akw}. The color map represents the TEBD results for open chains of sizes up to $L=280a$ and the black dashed lines are calculated within mean-field theory.
The overall qualitative agreement is rather good. 
The single-particle spectra in Figs.~\ref{fig:akw} (a),~(c), and (e) correspond exactly to the respective low-energy spectra of the noninteracting model at half filling, and the gapped nature of all phases is evident.
The flat bands in Figs.~\ref{fig:akw}(c) arise because there is no momentum dependence in a band in the parameter point $g=w=\Delta$~(Eq.~\ref{bulgSpec}).

Turning on strong interactions $U=1$ (right column of Fig.~\ref{fig:akw}) results in breaking of the particle-hole symmetry, seen in a corresponding shift of the spectral weight towards positive energies. 
Nevertheless,  the mean-field Hamiltonian remains $\mc P$ symmetric [Eq.~(\ref{PSym})], because it captures only the single-particle physics and retains the symmetry between the positive and negative states.
The band crossings of Figs.~\ref{fig:akw}(c) and (e) are lost, with a subgap opening at small momentum. 
Increasing $U$ leads to a decreasing of the bulk gap, cf. Fig.~\ref{fig:large-U}, such that it quickly becomes too small to observe with our energy resolution. Inside the topological superconducting phase Fig.~\ref{fig:akw}(d), the gap is not located at $k=0$, but moves at some small momenta which depend on $g$ and $\Delta$.
Also noteworthy is the appearance of a faint shoulder in $A(k,\omega)$ for $|k|$  close to $\pi$ in Fig.~\ref{fig:akw}(d) at $\omega\approx3$, which may signal the presence of a scattering continuum.
Unsurprisingly, a local probe in the bulk, such as $A(k,\omega)$, does not distinguish between phases with different topological character.

\subsubsection{Local spectral functions}

\begin{figure}[t]
\centering
\includegraphics[width=\columnwidth]{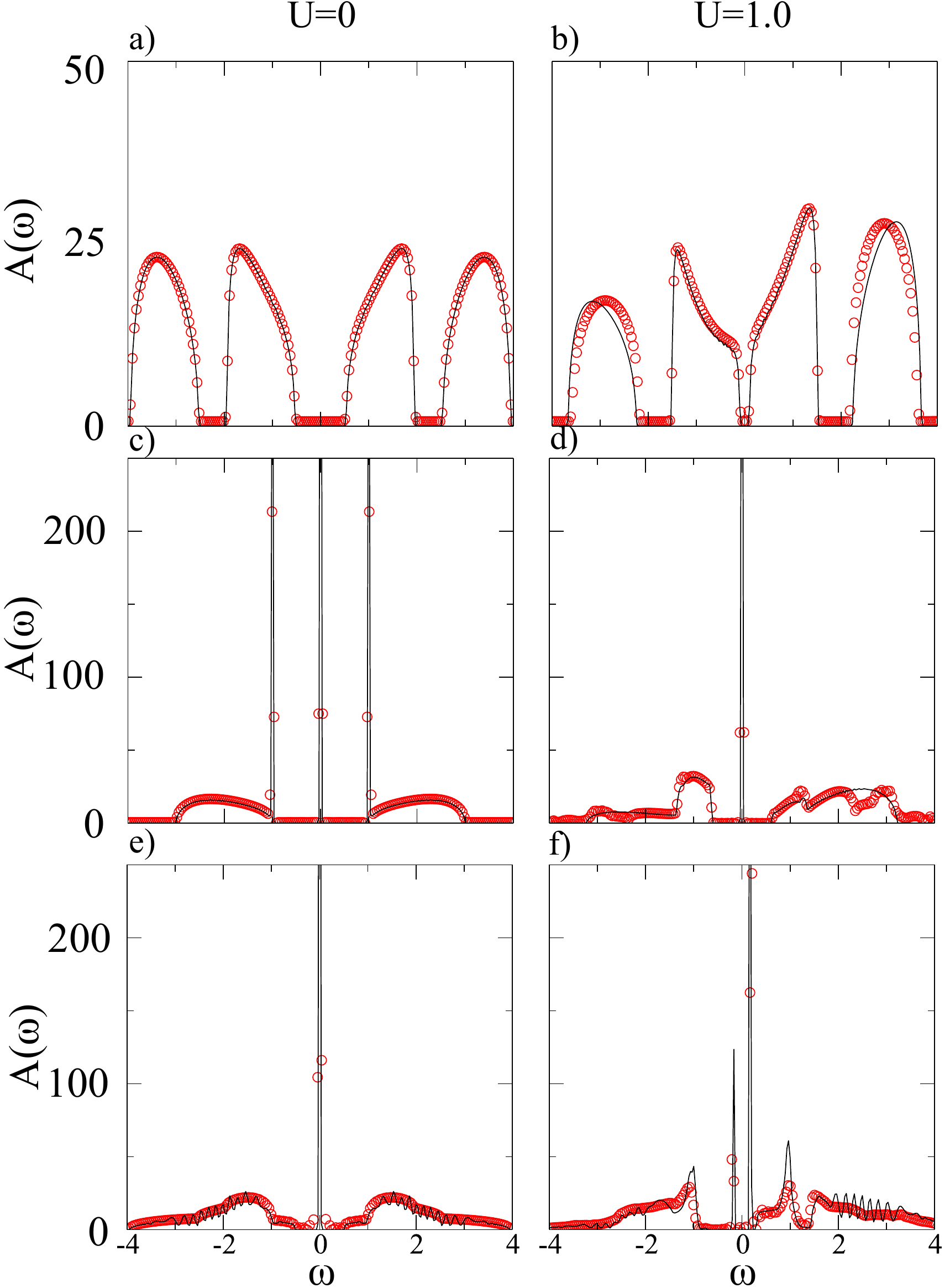} 
\caption{
(Color online)
Local spectral function $A_1(\omega)$ at the edge of an open wire for $U=0$  (left column) and $U=1$ (right column). 
(a), (b) The trivial phase at $g^2=0.5$ and $\Delta^2=5$ has no quasiparticle peak at zero energy.
(c), (d) A Majorana edge fermion in the topological phase at $g^2=\Delta^2=1$ is seen as a sharp peak at zero energy, which survives finite interactions.
(e), (f) The superconducting phase at  $g^2=5$ and $\Delta^2=0.5$. (e) The presence of zero-energy states is reflected in the spectral function peak at zero energy. (f) The CBS are not robust to interactions $U$, which removes them from zero energy, leading to a unique ground state.
Each subfigure combines TEBD (red open points) with mean-field results (lines),
scaled by an overall factor.
}
\label{fig:aw}
\end{figure}

We now turn our attention to the edge physics. 
In order to look at edge properties, we calculate the \emph{local} spectral function $A_1(\omega)=A(i=1,\omega)$ for open chains of size $L=140a$ for the same parameters as in Fig.~\ref{fig:akw}. The edge spectral function is represented in Fig.~\ref{fig:aw} using both mean-field results and DMRG approaches. Once again, the qualitative agreement between the two methods is rather good.

At the mean-field level, the Hamiltonian for the open chain is still described by Eq.~(\ref{meanFieldK}). However, the order parameters ($\rho,m,\eta,\delta$) on the open chain are determined from the Hartree-Fock calculation on the periodic system. The local density of states at site $i$ reads as
\begin{equation}
A_i(\omega)=\sum_\alpha |\langle i|\psi_\alpha\rangle|^2\delta(\omega-\epsilon_\alpha).
\end{equation}
The sum is carried over all eigenstates $|\psi_\alpha\rangle$ with respective eigenenergies $\epsilon_\alpha$. The Dirac delta function is numerically approximated  using Gaussians of width inversely proportional to the system size.

In the context of DMRG, the local spectral functions $A_i(\omega)$ are computed by adding a fermion (or removing) to an edge of the chain and then calculating only equal-space correlations, i.e., $i=i_0=1$ in Eqs.~(\ref{eq:GFs}) and (\ref{eq:GFs-2}).

The signatures of the trivial phase are not affected much by the presence of interactions, [cf. Figs.~\ref{fig:aw}(a) and~\ref{fig:aw}(b)].  The absence of zero-energy quasiparticle peaks remains, but the energy gap decreases with interactions.

More interestingly, the topological phase shows a clear signature of the Majorana zero-energy mode for both $U=0$ and $1$ in Figs.~\ref{fig:aw}(c) and~\ref{fig:aw}(d). This can  be interpreted as a direct observation of the Majorana bound state.
Without interactions, this mode is well separated by a symmetric gap of width $\omega=1$  from the remainder of the excitations.
The overall structure survives the addition of interactions, reflecting once again the robustness of the Majorana states.
The survival of the zero-energy MBS peak was also recently observed in the Kitaev chain subject to nearest-neighbor repulsive interactions.\cite{Thomale2013}

The edge CBS at large $g$ are also clearly signalled by the presence of a zero-energy peak in the noninteracting regime [Fig.~\ref{fig:aw}(e)].
In order to perform this simulation, the ground-state MPS is biased to one of the four degenerate states by applying a small pinning potential to an edge of the chain. 
Turning interactions to $U=1$ splits this peak into two peaks at nonzero energies, showing that the CBS have moved away from zero energy, as described in the preceding sections.

\section{Creutz-Hubbard model}
\label{sec:CH}
In this section, we focus on the original Creutz model without superconductivity, but with Hubbard interactions
\begin{equation}
H=H_{\rm C}+U\sum_j n_{j\up}n_{j\down}.
\end{equation}
In the noninteracting Creutz model, there are two insulating phases meeting at a critical gapless point $|g/w|=1$ (Fig.~\ref{fig:phdiag}). In contrast, at finite $U$, there is an extended phase developing around $|g/w|=1$ along the nonsuperconducting $\Delta=0$ line (Fig.~\ref{fig:MFDMRG}). The present section seeks to elucidate the nature of this phase.

Close to the Dirac point ($|g/w|=1$), adding a small interaction leads to moving the chemical potential from half filling into the bulk bands. Therefore, the system develops a metallic phase.
However, when the Creutz model is in an insulating phase, small interactions with respect to this gap, keep the chemical potential inside the gap and leave the system insulating. This explanation accounts for the observed increase in the metallic phase with the interaction strength, near the original band touching. 

\begin{figure}[t]
\includegraphics[width=\columnwidth]{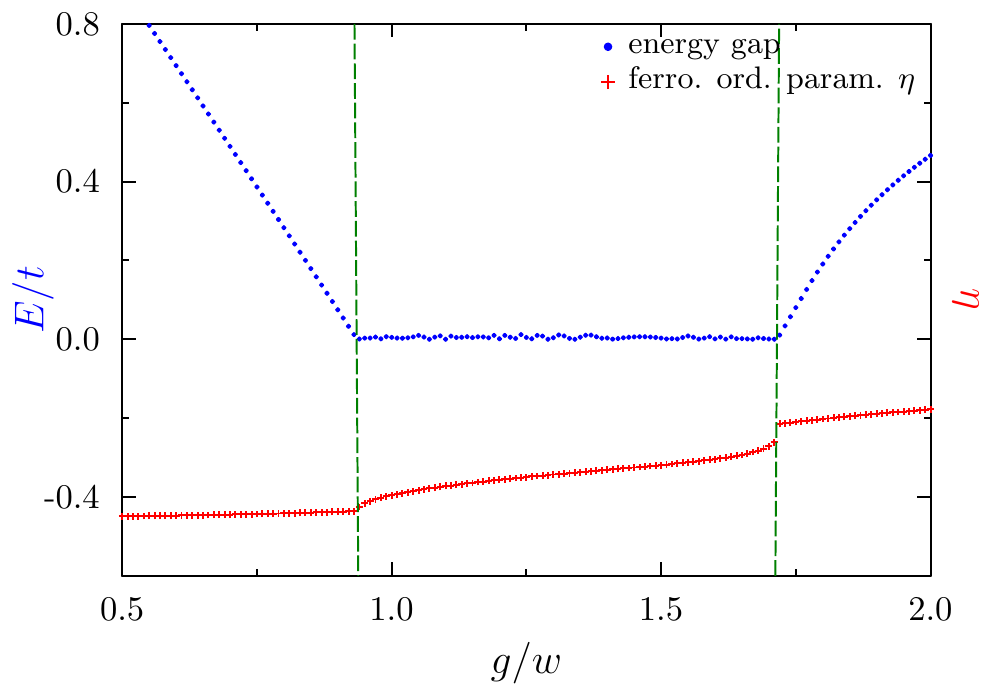}
\caption{(Color online) Energy gap for the interacting Creutz model at $U/t=1$ in mean-field theory (blue dots). Ferromagnetic order parameter $\eta$ in the $x$ direction (red crosses; in dimensionless units). Phase transitions are denoted by dashed green lines. The closing of the gap in the intermediary phase marks the metallic phase. The wire length is $L=500a$.}
\label{fig:gap}
\end{figure}

At the mean-field level, we consider explicitly in this case possible antiferromagnetic ordering in $x$ and, respectively, $z$ directions
\begin{equation}
\eta^a=(-1)^j\avg{c^\dag_{j\up}c_{j\down}},\quad
m^a=(-1)^j\avg{n_{j\up}-n_{j\down}}.
\end{equation}
However, both mean-field and DMRG indicate that there is only ferromagnetism in the $x$ direction developing for low to moderate $U$ (and vanishing ferromagnetic $m$, and antiferromagnetic $m^a$, $\eta^a$). Keeping the formalism where the degrees of freedom are doubled, the metallic region is identified with a closing of an energy gap. On Fig.~\ref{fig:gap} it is shown the system becoming metallic at precisely the predicted points on the phase diagram (Fig.~\ref{fig:MFDMRG}).

\begin{figure}
\includegraphics[width=4cm]{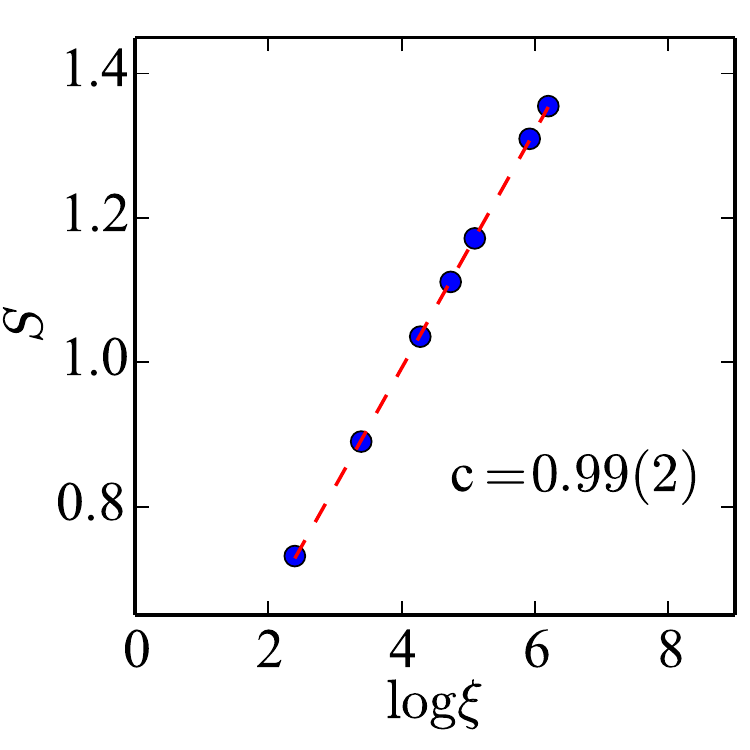}
\caption{(Color online) Finite-entanglement scaling of DMRG results for the interacting Creutz model at $g^2/w^2=2$ and $U/t=1$. The central charge $c\simeq 1$ matches that of a normal metallic phase.
}
\label{fig:metallic-dmrg}
\end{figure}

This picture is fully supported by DMRG results (cf. Fig.~\ref{fig:metallic-dmrg} for a representative case $g^2/w^2=2$ and $U/t=1$). The system is in a gapless metallic phase, with a central charge $c\simeq 1$ as for free fermions. Superconducting $s$-wave correlations are observed to decay exponentially fast with distance.
The interaction $U$ reinforces the $w$ term. For positive $w$, $\eta$ is negative such that the renormalized on-site coupling $\bar w=w-U\eta$ is larger than the bare $w$.
Nevertheless, the ``spin-orbit'' term $g(>0)$ competes with the ferromagnetic ordering $\eta$. It tends to delocalize the electrons and to spin-flip, thus leading to a reduction of the $\eta$ polarization at higher values of $|g|$ (Fig.~\ref{fig:gap}).

At finite superconducting pairing $\Delta$, the metallic region is immediately gapped out. Consequently, Majorana fermions are hosted in the proximity-induced gap, leading to the topological superconducting region in Fig.~\ref{fig:MFDMRG}.

\section{Conclusions}
\label{sec:conc}

This paper demonstrates that insulators hosting fractionally charged states (chiral bound states or CBS) continue to exhibit solitonic modes under the addition of proximity-induced superconducting pairing. It also explains in detail how the CBS can be transformed into Majorana bound states (MBS) when the pairing exceeds a certain threshold fixed by the gap of the insulator. Building on two representative examples, the Creutz model and the Su-Schrieffer-Heeger model (treated in the Appendix), we propose the following general mechanism for one-dimensional topological insulators in the BDI class. 

For moderate pairing, the CBS remained pinned at zero energy owing to the protection of chiral symmetries. This is reflected in a fourfold-degenerate many-body ground state for a finite wire. Indeed, the two zero-energy end states of the wire can be either occupied or empty without changing the total energy of the electronic system. However, breaking the chiral symmetries, pushes the CBS at finite energy, thereby leading to a unique ground state (the chemical potential is fixed at zero energy) because the localized end states are then either completely filled or completely empty.

At stronger superconducting pairing, larger than the initial insulating gap, the Creutz, and the Su-Schrieffer-Heeger insulators are driven in a phase supporting Majorana fermions. In the absence of interaction, we have investigated this transition directly by finding the wave functions of the zero mode pinned at the interface between a topological and a trivial semi-infinite phase. This transition can be also seen in a halving of the many-body ground-state degeneracy for a finite wire: the two end states become then two spatially separated Majoranas forming a single ordinary fermion (topological quantum bit).

The behavior of the CBS and MBS is also considered under the effects of repulsive interactions. In the Creutz model, the interactions were shown to break the chiral symmetry, leading to a removal of the CBS from zero energy and a unique ground state. 
In contrast, the Majorana fermions are rather robust, such that the parameter regime for the existence of Majorana fermions increases, albeit the bulk gap is reduced by interactions.

Aside from the edge-state physics, we have also studied the bulk phase diagram of the Creutz-Majorana-Hubbard model using a combination of self-consistent Hartree-Fock theory and extensive DMRG simulations (cf.\ Fig.~\ref{fig:MFDMRG}), the agreement between both approaches being overall very good. The main feature is that increasing the Hubbard repulsion $U$ tends to expand stability domain of the topological Majorana superconductor, in agreement with the phenomenology obtained in other nanowire models.\cite{Stoudenmire2011} In contrast, at weak pairing $|\Delta/w|<1$ and small spin-orbit coupling $|g/w|<1$, the interactions lead to a decrease of the Majorana phase (cf.\ Fig.~\ref{fig:MFDMRG}). 
For large $|g|$, the model exhibits a topological phase which is protected by a combination of fermion parity and inversion symmetry.

Besides the  topological phases, the phase diagram (see Figure~\ref{fig:MFDMRG}) presents various other interesting bulk phases. For example, in the absence of superconductivity ($\Delta=0$), a metallic phase develops for a finite range of values (which extends upon raising the interaction $U$) of the spin-orbit parameter $g/w$ owing to the effect of interactions. This gapless phase was restricted to $|g/w|=1$ in the absence of interactions. It would be interesting to study the possibility of intrinsic superconductivity in such a phase.

The authors thank  E.~Orignac and S.~Huber for interesting discussions. D.~S. thanks P.~Simon for drawing his attention on the subject of fractionally charged solitons. J.~C.\ acknowledges support from EU/FP7 under contract TEMSSOC. D.~S. and J.~C. were supported by the French ANR through projects ISOTOP and MASH.

\appendix*
\section{Superconducting SSH model} 
\label{sec:sssh}
The physics of survival of solitonic modes into the superconducting phase and their subsequent transformation into Majorana fermions is more general. This section comments on the topological phases in another model from the BDI class, the Su-Schrieffer-Heeger (SSH) model.\cite{Su1979}

\begin{figure}[t]
\centering
\includegraphics[width=0.85\columnwidth]{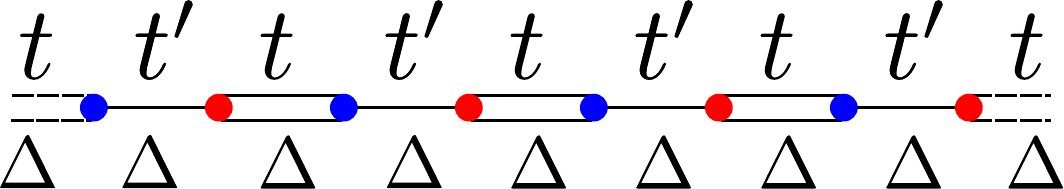}
\caption{(Color online) Superconducting SSH model. The red A sites and the blue B sites are connected by alternating hopping terms leading to a dimer structure. A homogeneous pairing $\Delta$ couples near-neighbor sites.}
\label{fig:sssh}
\end{figure}

\begin{figure}[t]
\includegraphics[width=0.75\columnwidth]{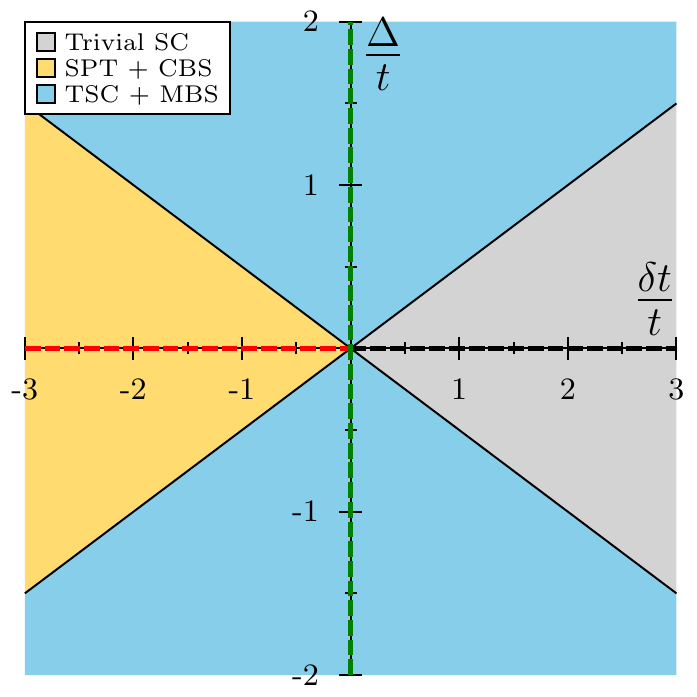}
\caption{(Color online) Phase diagram of the superconducting SSH model. The original SSH is at $\Delta/t=0$ axis, marked by the dashed line. Dashed red line is the region with fractional charge solitons, and in black the trivial insulating region. Under the addition of superconducting $p$-wave pairing, the solitonic (CBS) region (represented in yellow) survives until the bulk gap closes. The region hosting CBS is an inversion symmetry-protected phase (SPT) in the bulk. At large $\Delta/t$, the system is in a topologically nontrivial region with one Majorana fermion trapped at the ends. The gray region is the trivially gapped state. The $y$ axis (green dashed line) represents the Kitaev chain model at zero chemical potential,\cite{Kitaev2001} in a nontrivial regime, for $\Delta\ne 0$.}
\label{fig:sshDiag}
\end{figure}

The most simple superconducting SSH model is represented by a dimerized chain with a cell containing two atoms A and B:
\begin{equation}
H=\sum_i tc^\dag_{Ai}c^{\phantom\dag}_{Bi}+t'c^\dag_{Bi}c^{\phantom\dag}_{Ai+1}
+\Delta c^\dag_{Ai}c^\dag_{Bi}+\Delta c^\dag_{Bi}c^\dag_{Ai+1}+\hc.
\end{equation}
The alternating hopping strengths represent the ``single bond'', $t'$, and the ``double bond'', $t$, of the polyacetylene chain. This model captures only the fermionic degrees of freedom from the Hamiltonian in Ref.~\onlinecite{Su1979}, neglecting the bosonic degrees of freedom due to lattice distortion. Crucially, the model is enriched here by the addition of a homogeneous superconducting order parameter $\Delta$ which couples near-neighbor sites (Fig.~\ref{fig:sssh}).

Let us consider first a finite one-dimensional wire in the absence of superconductivity ($\Delta=0$) with an integer number of cells ($A-B$) and positive real values for $t$ and $t'$. If the inter-cell hopping strengths are larger than the intra-cell hopping terms, $\delta t=t-t'<0$, then a zero-energy state is bound at the edge. 

As in the CM model, the solitonic modes survive at zero energy under the addition of a finite superconducting pairing. In this case, they are protected by the insulating gap induced by the Peierls instability $\propto \delta t$.
The general mechanism for their survival depends on the relative magnitude between the insulating gap and the superconducting gap. The topology of the Hamiltonian is characterized using Eq.~(\ref{MajNo}) and the resulting phase diagram is represented in Fig.~\ref{fig:sshDiag}.

For small superconducting pairing $|\Delta|<|\delta t/2|$, the system has two phases without Majorana fermions. These are two distinct phases: one is an inversion symmetry-protected phase (only in the bulk), and with solitonic modes (CBS) that survive at finite $\Delta$ until entering the Majorana region ($\delta t<0$); the other is a trivial phase devoid of midgap states ($\delta t>0$).

For large superconducting pairing $|\Delta|>|\delta t/2|$, the system is in a topologically nontrivial phase which supports Majorana fermions at the ends. Note that Kitaev chain model\cite{Kitaev2001} at zero chemical potential is recovered by making the system uniform in normal hopping, $\delta t=0$.  In this case, any finite $\Delta$ yields one Majorana bound state at the edge.

In the presence of repulsive interactions, we expect to see a similar removal of the ground-state degeneracy in the CBS phase. This was already investigated in the absence of superconductivity, revealing a split in energy between a charged soliton and the neutral solitons.\cite{Kivelson1982} The addition of superconducting pairing should qualitatively change these results only by allowing a Majorana phase robust to interactions.

In this section, we have shown how another BDI system supporting fractionally charge modes can be driven in a Majorana phases under the effect of superconducting pairing.
In a straight parallel to the Creutz model, the solitonic modes are protected by a chiral symmetry and remain at zero energy in a wide region of parameters, before entering a Majorana phase. The condition to see the transition to the Majorana phase comes to a competition between the insulating gap and the induced superconducting pairing.

\bibliographystyle{apsrev4-1}
\bibliography{bibl}
\end{document}